\documentclass[twocolumn]{aastex631}
\usepackage{float}
\usepackage[utf8]{inputenc}
\usepackage{amsmath}
\usepackage{makecell}

\begin{document}

\title{The Quasi-Radial Field-line Tracing (QRaFT): an Adaptive Segmentation of the Open-Flux Solar Corona}

\correspondingauthor{Vadim M. Uritsky}
\email{vadim.uritsky@nasa.gov}

\author[0000-0002-5871-6605]{Vadim M. Uritsky}
\affiliation{The Catholic University of America \\
 620 Michigan Avenue NE \\ 
 Washington, DC 20064, USA}
\affiliation{NASA Goddard Space Flight Center \\
Code 670, Greenbelt, MD 20771, USA}

\author[0000-0002-4992-180X]{Christopher E. Rura}
\affiliation{The Catholic University of America \\
 620 Michigan Avenue NE \\ 
 Washington, DC 20064, USA}
\affiliation{NASA Goddard Space Flight Center \\
Code 670, Greenbelt, MD 20771, USA}

\author[0000-0003-1759-4354]{Cooper Downs}
\affiliation{Predictive Science Inc. \\
9990 Mesa Rim Road, Suite 170, San Diego, CA 92121, USA}

\author[0000-0001-9498-460X]{Shaela I. Jones}
\affiliation{The Catholic University of America \\
 620 Michigan Avenue NE \\ 
 Washington, DC 20064, USA}
\affiliation{NASA Goddard Space Flight Center \\
Code 670, Greenbelt, MD 20771, USA}

\author[0000-0001-9326-3448]{Charles Nickolos Arge}
\affiliation{NASA Goddard Space Flight Center \\
Code 670, Greenbelt, MD 20771, USA}

\author[0000-0001-5207-9628]{Nathalia Alzate}
\affiliation{ADNET Systems, Inc. \\
Greenbelt MD 20771, USA}
\affiliation{NASA Goddard Space Flight Center \\
Code 670, Greenbelt, MD 20771, USA}

\begin{abstract}

Optical observations of solar corona provide key information on its magnetic geometry. The large-scale open field of the corona plays an important role in shaping the ambient solar wind and constraining the propagation dynamics of the embedded structures, such as interplanetary coronal mass ejections. Rigorous analysis of the open-flux coronal regions based on coronagraph images can be quite challenging because of the depleted plasma density resulting in low signal-to-noise ratios. In this paper, we present an in-depth description of a new image segmentation methodology, the Quasi-Radial Field-line Tracing (QRaFT), enabling a detection of field-aligned optical coronal features approximating the orientation of the steady-state open magnetic field. The methodology is tested using synthetic coronagraph images generated by a three-dimensional magnetohydrodynamic model. The results of the numerical tests indicate that the extracted optical features are aligned within $\sim 4-7$ degrees with the local magnetic field in the underlying numerical solution. We also demonstrate the performance of the method on real-life coronal images obtained from a space-borne coronagraph and a ground-based camera. We argue that QRaFT outputs contain valuable empirical information about the global steady-state morphology of the corona which could help improving the accuracy of coronal and solar wind models and space weather forecasts.

\end{abstract}

\section{Introduction}
\label{sec:intro}
The magnetic field of the coupled solar photosphere and corona is among the most important factors that define both the steady-state and the eruptive plasma evolution in the surrounding heliosphere \citep{beedle2022}. The coronal magnetic field is often considered to be the primary free energy reservoir for the processes in the lower corona \citep{solanki2006}. It also shapes the geometry and connectivity of the corona with the heliosphere \citep{arge2002}, defines the pathways of bulk plasma transport \citep[e.g.][and refs therein]{filho2024}, and controls the stability of the coronal plasma at different scales \citep{uritsky2013, uritsky2022a, klimchuk2023}. 

The inherent uncertainty of remote observations of the solar magnetic field limits the accuracy of global heliospheric models used for space weather forecasts. As of now, the line-of-sight photospheric magnetic field continues to be the most reliable observable used to constrain operational solar wind models. Thus, for instance, the widely used Wang-Sheeley-Arge (WSA) empirical model is driven by systematically updated photospheric field synoptic maps \citep{wang1990, arge2000, arge2004}. The WSA solutions provide the solar wind context for coronal mass ejections (CME) events which can be analyzed in more detail using magnetohydrodynamic (MHD) modeling frameworks \citep{odstrcil2004, odstrcil2005, merkin2016, provornikova2024}. The inner portion of WSA uses the magnetostatic potential field source surface (PFSS) model \citep{altschuler1969, wang1992} to extrapolate the coronal field from the inner photosperic boundary at $R_S$ to the outer source surface boundary at about $2.5 R_S$. The outer portion is the Schatten current sheet model \citep{schatten1971}  enabling a more realistic description of the outer corona. The two parts of the WSA model are coupled through the radial component of the magnetic field, making the uncertainty in the open magnetic flux one of the main factors limiting the accuracy of the WSA model, as well as MHD solar wind models relying on WSA solutions. 

The magnitude of the total unsigned open flux predicted by most of the available models appears to be inconsistent with in situ solar wind observations \citep{riley2019, wallace2019, badman2021,linker2017, linker2021, wang2022}, suggesting that the open magnetic field of the corona is not adequately represented. Some of this disagreement was attributed to a variety of sources, including an underestimated polar flux and the lack of direct observation on the far side of the Sun \citep[see e.g.][and refs therein]{linker2017}, problems with photospheric magnetic maps \citep{posner2021}, and transient processes at the boundaries of coronal holes \citep{arge2024}. 

The errors in the description of the open-field magnetic geometry below the source surface are amplified with the heliocentric distance due to the field-line expansion and nonlinear effects, leading to a less reliable description of the ambient solar wind flow \citep{wallace2020, jones2022}, and, as a consequence, to inaccurate forecasts of CMEs and other types of embedded propagating disturbances, especially when they possess a complex magnetic structure and actively interact with the environment \citep[see e.g.][]{pudovkin1979, zurbuchen2006, mays2020, odstrcil2023}.

Much of the uncertainty in the simulated open coronal flux is known to be caused by the inherent limitations of input data used by the models, i.e. global photospheric magnetic field maps. \citet{jones2016, jones2017,jones2020} have proposed an innovative approach for improving photospheric magnetograms by using additional empirical constraints obtained from coronagraph images. They have shown that the structure of the observed white-light corona can provide valuable information about the quality of the synoptic magnetograms used by the PFSS. They have also demonstrated that the discrepancy between the observed and simulated coronal geometry can be reduced by optimizing the photospheric boundary condition \citep{jones2020}. The optimization scheme developed in these series of papers utilized the outputs of a coronal image segmentation algorithm which subsequently evolved into QRaFT -- a sophisticated image processing methodology implemented as a stand-alone data analysis package available to the community \citep{uritsky2022,uritsky2024}. 

The goal of this paper is to provide an exhaustive description of the mathematical methods and processing steps involved in QRaFT, to evaluate its uncertainties, and to demonstrate examples of its application to coronagraph images. In a companion paper \citep{rura2025}, we give a more detailed account of the validation and testing of the QRaFT method using synthetic coronal images derived from a global coronal MHD model. 

The paper is organized as follows. In Section \ref{sec:QRaFT}, we present the main processing algorithms, control parameters, and data metrics used by QRaFT. Section \ref{sec:performance} is dedicated to QRaFT performance tests on several types of images, including  synthetic coronal images constructed using an MHD model making it possible to compare image segmentation results with the ground-truth magnetic field (Section \ref{sec:performance_MAS}), a coronal image obtained from the space-borne COR1 coronagraph onboard STEREO mission (Section \ref{sec:performance_COR1}), and a ground-based coronal image taken during a total solar eclipse (Section \ref{sec:performance_TSE}). Section \ref{sec:conclusions} summarizes the paper and discusses possible future applications of QRaFT.

%********* (see PSI_results.docx file ***********
\section{Detecting quasi-radial coronal features}
\label{sec:QRaFT}

\subsection{General considerations}
\label{sec:QRaFT_general}

The plasma of the lower corona is constrained by the magnetic field generated in the convection zone. The motions of the particles and the associated thermodynamic properties along and across the zero-order magnetic field are significantly different. The diffusion of the coronal thermal energy is much more efficient in the magnetic field direction \citep{ye2020,navarro2022}, with the predicted ratio of the perpendicular to parallel thermal conduction being of the order of $10^{-10}$ or smaller \citep{vanhoven1984}. As a result, mass and energy transport in the corona primarily occur along the magnetic field lines. The combination of the greatly suppressed cross-field thermal conduction with the low thermal to magnetic pressure ratio and the frozen-in flux condition leads to a nearly one-dimensional behavior in which the plasma material and thermal energy are channeled along quasi-rigid, thermally insulated magnetic flux tubes \citep{klimchuk2006,klimchuk2015}. As a result, the field-aligned density and energy gradients tend to be inherently transient, and their relaxation time is many orders of magnitude smaller than that of cross-field gradients which dominate steady-state regimes. 

This anisotropy plays a crucial role in remote coronal observations justifying the assumption that long-living optical gradients in either closed or open corona represent the local orientation of the coronal magnetic field lines. This assumption has been used by a large number of image tracing algorithms designed by the solar community. Most of these algorithms belong to one of the two categories: (1) closed-field tracing of solar active regions (ARs) in the extreme ultraviolet corona \citep[e.g.][]{aschwanden2008, aschwanden2010, aschwanden2013, malanushenko2014} or (2) open-field tracing of white-light ground-based total solar eclipse (TSE) images \citep[e.g.][]{durak2009, boe2020, bemporad2020, bemporad2023}. Both groups of image segmentation methods rely on the high signal-to-noise ratio (SNR) of the processed images ensured by either the strong optical emission of the hot and dense AR loops \citep[see e.g.][and refs therein]{kniezewski2024, uritsky2024a} or the unique observing conditions associated with the TSE photography \citep{bemporad2023}. 

To our knowledge, no method other than QRaFT is currently available for tracing open-field structures in routinely collected non-TSE coronal images which could be used in operational space weather forecasts. The QRaFT methodology has been developed to fill this gap. The core idea of QRaFT is to improve the SNR of the  azimuthal (parallel to the Sun's surface) optical gradients used for field-line tracing at the expense of radial gradients which carry little information about the open-field magnetic structures. Exploiting this trade-off enables us to preserve the fine structure of the open-flux corona and to quantify its plane of sky (POS) geometry. 

The first applications of the QRaFT technique have demonstrated its ability to recover physically relevant coronal morphologies \citep{jones2020}, which can be used to improve the quality of synoptic magnetograms and the accuracy of potential-field extrapolation models \citep{jones2016, jones2017}. Below we provide a full description of the publicly available release of the QRaFT package \citep{uritsky2024} to facilitate its wider application to various types of coronal imaging data and models. 

The rest of this section outlines the main processing steps involved in QRaFT, which include the initial image preparation enhancing the small-scale azimuthal structures communicating the information about the magnetic field geometry in the open-field coronal regions, the anisotropic noise reduction preserving the fine structure of quasi-radial features, the tracing and interpolation routines enabling a rigorous mathematical representation of the detected features, as well as filtering and validation routines performing automated quality checks of the tracing results. 

\subsection{Image enhancement}
\label{sec:QRaFT_enh}

The initial image enhancement begins with the removal of the large-scale radial trend. Typically, the available coronagraph data products undergo radial detrending as a part of standard calibration and preprocessing procedures, but there is a substantial residual brightness trend that needs to be removed to enable a consistent feature detection across the range of solar distances covered by the instrument. 

We first compute the averaged trend function $I_{tr}(r)$ in rectangular coordinates by averaging over groups of image pixels characterized by the same distance $r$ from the solar disk center:

\begin{eqnarray}
\label{eq:trend}
I_{tr}(r) & = & \langle I(x,y) \rangle_{\Omega(r)},  \\
\Omega(r) & = & \left\{ (x, y) \,|\, (x-x_0)^2+(y-y_0)^2=r^2 \right\} 
\end{eqnarray}

Here, $I(x,y)$ is the brightness of the original image at the Cartesian location $(x,y)$, $x_0$ and $y_0$ are the coordinates of the disk center, and $\Omega$ is the set of pixels constituting a ring around the Sun characterizing by a given $r$ (within the image resolution uncertainty).  

Subtracting the trend function from the original image results in its detrended version, in which the large-scale radial decay of the brightness is eliminated: 

\begin{equation}
\label{eq:detr}
I_{detr}(x,y) = I(x,y) - I_{tr} \left( \sqrt{(x-x_0)^2+(y-y_0)^2}\right)
\end{equation}

Next, the detrended image $I_{detr}(x,y)$ undergoes is transformed into plane polar coordinates $(\phi, \rho)$ representing the POS counterclockwise azimuth angle and the radial position of the pixel, correspondingly:

\begin{equation}
\label{eq:polar}
I_{detr}(x,y) \rightarrow I(\phi, \rho)
\end{equation}

Since it is important to preserve the fine structure in the transformed image, the bins in the $\phi$ and $\rho$ directions are chosen to be barely enough to produce a smooth rectangular-to-polar mapping, which may result in a small number of isolated empty bins, typically at the largest radial distances. These bins are filled with the local average intensity values of the non-empty nearest-neighbor pixels.

The obtained polar-coordinate array $I(\phi,\rho)$ undergoes a linear low-pass filtering intended to reduce the noise level and to ensure a more robust finite differencing result at the subsequent step: 

\begin{equation}
\label{eq:smooth}
I_s(\phi, \rho) = \\
\frac{1}{\Delta\phi_s \Delta\rho_s} \int_{\rho-\Delta\rho_s/2}^{\rho+\Delta\rho_s/2} \!\!\! \int_{\phi-\Delta\phi_s/2}^{\phi+\Delta\phi_s/2} \!\!\!\! I(\phi', \rho')\, d\phi'\,d\rho'
\end{equation}
where $\Delta\phi_s$ and $\Delta\rho_s$ are the smoothing scales in respectively $\phi$ and $\rho$ directions. The smoothing in the azimuthal direction uses the periodic boundary condition $I(\phi=0,\rho) = I(\phi=2\pi, \rho)$. The smoothed array $I_s(\phi, \rho)$ is used to compute the unsigned second-order azimuthal derivative estimated based on the second-order centered difference approximation:

\begin{equation}
\label{eq:diff}
I^{''}(\phi, \rho) = | I_s(\phi-\Delta \phi/2, \rho) + I_s(\phi+\Delta \phi/2, \rho) - 2 I_s(\phi, \rho) |,
\end{equation}
in which $\Delta \phi$ is the characteristic scale of the azimuthal differencing. The unsigned second-order azimuthal derivative is an efficient indicator of quasi-radial structures with both simple and complex morphology as shown below.

%-----------------------------------
\begin{figure*}
\begin{center}
\includegraphics[width=15.0 cm]{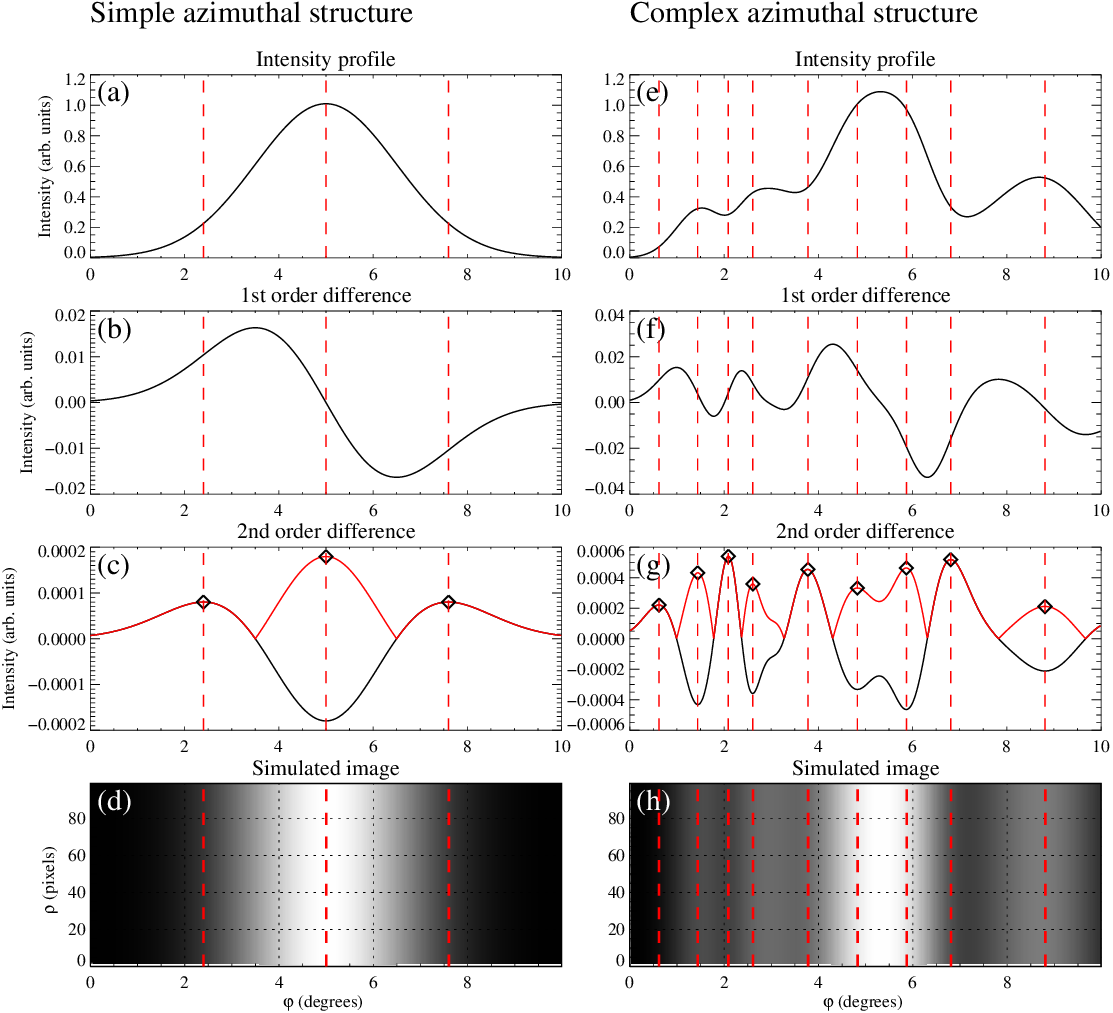}
\caption{\label{fig_diff} Segmenting simulated quasi-radial coronal structures using unsigned second-order azimuthal differencing. Left column: a simple model of an azimuthal intensity profile represented by a single Gaussian function. Right column: A more complex model composed of several randomly generated Gaussian peaks. Panels (a) and (e): processed intensity profiles; panels (b) and (f): first-order azimuthal differences; panels (c) and (g): signed (black) and unsigned (red) second-order azimuthal differences, panels (d) and (h): simulated POS intensity variability corresponding to the studied profiles. Red dashed vertical lines show local maxima of the unsigned second-order difference (shown with diamonds on panels (c) and (g)). These lines are aligned with the large-scale density gradients which are expected to be constrained by the coronal magnetic field, enabling an image-based field-line tracing.}
\end{center}
\end{figure*}
%-----------------------------------

In order to further enhance the azimuthal structure in the differenced array $I^{''}$, we next eliminate its large-scale azimuthal trend. This goal is achieved by dividing $I^{''}$ by the trend array $I_{tr}^{''}$ containing the smoothed azimuthal profile of the image: 

\begin{eqnarray}
\label{eq:enh}
I_{tr}^{''}(\phi) & = & \frac{1}{\Delta\phi_d} \int_{\phi-\Delta\phi_d/2}^{\phi+\Delta\phi_d/2} \langle I^{''}(\phi',\rho) \rangle_{\rho} \, d\phi', \\ 
I_{enh}(\phi,\rho) & = & \frac{I^{''}(\phi, \rho)}{I_{tr}^{''}(\phi)}, %\,\,\, \forall \, \rho
\end{eqnarray}
where $\langle . \rangle_\rho$ denotes averaging over all $\rho$ positions and $\Delta\phi_d$ is the characteristic scale of the azimuthal detrending. A proper regularization is applied to avoid the divergence of $I_{enh}$ at the location of small $I_{tr}^{''}$.

The image processing methods described in this section involve a number of tunable control parameters ($\Delta\phi_s$, $\Delta\rho_s$, $\Delta\phi$, $\Delta\phi_d$) which can be optimized to obtain the best possible image enhancement results for a particular type of coronal images. Our tests have shown that the optimization of these parameters depends on the image resolution and noise level and can also be affected by the outward boundary of the studied coronal region. The preprocessing steps described above are illustrated in Section \ref{sec:performance}, with the optimized processing parameters listed in Table \ref{tab:keywords}.

\subsection{Tracing image features}
\label{sec:QRaFT_tracing}

The enhanced polar-coordinate image array $I_{enh}$ contains a detailed information about the location of radially-aligned coronal structures, as illustrated by two numerical tests shown in Fig. \ref{fig_diff}. In the first test, the azimuthal profile is simulated by a single Gaussian function. The unsigned second-order difference of the azimuthal profile contains three easily identifiable peaks which can be used to trace the structure by applying two properly adjusted detection thresholds, one for the crest and one for the wings of the structure. The second test is based on a more complex mathematical model including several randomly positioned Gaussian functions of various width. In this case, a larger set of thresholds would be needed to detect each of the features contained in the simulated profile. 
 
These examples demonstrate that it should be possible to detect multiple overlapping quasi-radial structures contained in the studied coronal image by applying a set of adaptively chosen detection thresholds to the enhanced image array $I_{enh}(\phi,\rho)$. The contiguous clusters of image pixels above each threshold can be used to identify the locations and spatial orientations of azimuthal image gradients characterized by significantly different optical intensities. This processing step  is implemented as follows.

We begin by defining a set $\{p_k\}$ of percentile probabilities, where $k=1,...,N_p$, used to calculate the percentile thresholds $I^{th}_k$: 

\begin{equation}
\label{eq:thresh}
\frac{\sum_{\phi,\rho} \mathcal{H}\left( I_{enh}(\phi, \rho) - I^{th}_k \right)}{N} = p_k.
\end{equation}

Here, $\mathcal{H}$ is the Heaviside step function returning a binary array whose elements equal 1 if the corresponding pixel of the enhanced image is above the threshold, and equal 0 otherwise $N$ is the total number of elements in $I_{enh}$. The criterion (\ref{eq:thresh}) ensures that the normalized occurrence rate of pixel values satisfying the condition $I_{enh} > I^{th}_k$ is approximately equal to the required probability $p_k$.

In addition to $\{p_k \}$, we also introduce a set $\{ \rho^{\min}_j \}$ of closest to the Sun $\rho$-positions including in the threshold-based detection, where $j = 1,...,N_{\rho}$. Each $\rho^{\min}_j$ represents an inner radial boundary above which the detection thresholds $I^{th}_k$ are applied and below which the pixels are ignored. By moving this boundary progressively upward and repeating the detection process, we refocus the algorithm on the increasingly higher, and optically weaker, coronal regions, while preventing the regions which are closer to the Sun from dominating the detection results due to their stronger signal. The sliding inner boundary mitigates the effects of the residual radial trend which may be present in the image after the inital detrending (\ref{eq:detr}). 

The detection algorithms loops through the whole set of percentile thresholds for every new inner boundary, which produces $N_p \times N_\rho$ individual detection runs. Each run outputs an array $\Lambda_{j,k}(\phi, \rho)$  filled with interger-valued pixel labels, ranging from 1 through $\lambda_{j,k}^{\max}$, which mark the locations of all contiguous pixel clusters satisfying the detection conditions:

\begin{equation}
\label{eq:lbl}
I_{enh}(\phi, \rho >\rho^{\min}_j) > I^{th}_k \,\,\, \rightarrow \,\,\, \Lambda_{j,k}(\phi, \rho) \in \{1, ... , \lambda^{\max}_{j,k}\},
\end{equation}
where the cluster labeling is performed using the {\textit {label\_region()}} function of the {\textit {IDL}} language. The maximum label value $\lambda_{j,k}^{\max}$ equals the total number of clusters found for a given combination of $I^{th}$ and $\rho^{\min}$. 

\begin{figure}
\begin{center}
\includegraphics[width=9 cm]{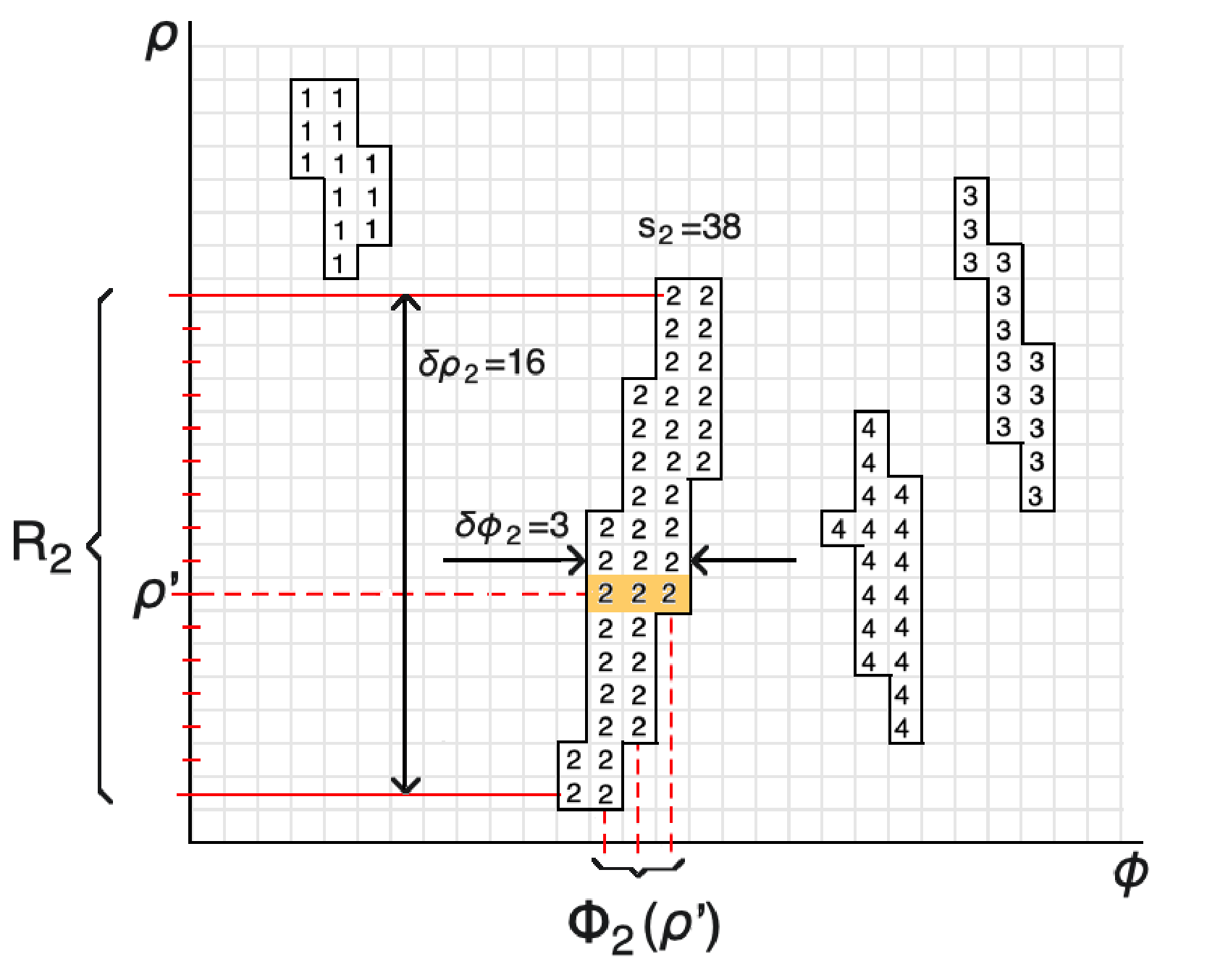}
\caption{\label{fig_clusters} Explanation of the geometric parameters of pixel clusters detected by QRaFT. The number assigned to each pixel represents the label $\lambda$ of the cluster to which it belongs as well as the subscript of its metrics defined in Section \ref{sec:QRaFT_metrics}. $R$ is the set of the radial positions involved in the cluster (\ref{eq:rho_set}),  $\Phi(\rho)$ is the two-level set of the azimuthal positions corresponding to each $\rho$ (\ref{eq:phi_set}), $\delta\rho$ is the radial extent of the cluster (\ref{eq:rho_size}),  $\delta\phi$ is the azimuthal thickness of the cluster (\ref{eq:phi_size}), and $s$ Is the the discrete area of the cluster  (\ref{eq:s_size}). }
\end{center}
\end{figure}

The cluster label maps $\Lambda_{j,k}(\phi, \rho)$ are next analyzed by the code to compose the set $\Phi_{\lambda}$ of $\phi$ positions for each $\rho$ position within a given cluster described by label $\lambda$, and the set $R_{\lambda}$ of all $\rho$ positions within that cluster (Fig. \ref{fig_clusters}):

\begin{eqnarray}
\label{eq:phi_set}
\Phi_\lambda (\rho) &=& \{ \phi \,|\, \Lambda(\phi,\rho) = \lambda  \}\\
%R_\lambda (\phi)= \{ \rho \,|\, \Lambda(\phi,\rho) = \lambda \}
\label{eq:rho_set}
R_\lambda &=& \{ \rho \,|\, (\exists \phi) [\Lambda(\phi,\rho) = \lambda ] \},
\end{eqnarray}
in which we omitted the indices $j$ and $k$ for brevity. 

%-----------------------------------
%-----------------------------------

At the next step, we form the array $\rho_{\lambda}$ of radial nodal points within the studied cluster, sorted in ascending order: 

\begin{equation}
\label{eq:rho_array}
\rho_{\lambda, i} = \min R_\lambda + i \, \Delta \rho,  
\end{equation}
where $\Delta\rho$ is the radial discretization bin size, $i = 0, ..., n_\lambda$ is the index of the radial position of the node, and $n_\lambda = (\max R_\lambda -\min R_\lambda)/\Delta \rho$ is the total number of such positions in a given cluster. For each $\rho_\lambda$, we compute the average azimuthal position of the cluster pixels:

\begin{equation}
\label{eq:phi_array}
\phi_{\lambda, i} = \langle \Phi_\lambda (\rho_{\lambda, i})  \rangle     
\end{equation}

The dependence of $\phi_\lambda$ on $\rho_\lambda$ defines an empirical line passing through the centroids of the cluster pixels located at the same radial distance. Due to the discretization noise, this line tends to exhibit a small-scale irregularity, which is especially pronounced when $s_\lambda$ is small. To reduce this irregularity, we apply the polynomial interpolation

\begin{equation}
\label{eq:phi_poly}
\phi_{\lambda, i}^* = \sum_{m=0}^M a_{\lambda,m} \, \rho_{\lambda, i}^m. 
\end{equation}

 Here, $M$ is the order of the polynomial, $a_{\lambda, m}$ are the polynomial coefficients of the $\lambda^{th}$ cluster, and $\phi_{\lambda, i}^*$ is the array of its interpolated azimuthal coordinates. The interpolation procedure is important since it improves the accuracy of the subsequent calculation of local orientation angles of the clusters. Approximating $\phi_\lambda$ for a given $\rho_\lambda$, rather than $\rho_\lambda$ for a given $\phi_\lambda$, is computationally more straightforward when analyzing quasi-radial coronal structures, since for such structures $\phi_\lambda(\rho_\lambda)$ is always a single-valued function while $\rho_\lambda(\phi_\lambda)$ is often not.

In the subsequent discussion, the interpolated lines describing the shapes of the detected pixel clusters are referred to as image {\it features}, and they represent the main product of the tracing algorithm implemented in QRaFT.

\subsection{Feature metrics}
\label{sec:QRaFT_metrics}

The constructed sets (\ref{eq:phi_set} - \ref{eq:rho_set}) of cluster coordinates enable quick determination of the basic parameters characterizing the geometry of the detected structures: the maximum azimuthal cluster thickness $\delta\phi_\lambda$, the radial cluster length $\delta\rho_\lambda$, and the discretized cluster area $s_\lambda$:

\begin{eqnarray}
\label{eq:phi_size}
\delta \phi_\lambda &=& \max_\rho | \Phi_\lambda(\rho) |, \\
\label{eq:rho_size}
\delta \rho_\lambda &=& \max R_\lambda - \min R_\lambda, \\   
\label{eq:s_size}
s_\lambda &=& |\{(\phi, \rho) \, | \, \Lambda(\phi, \rho) = \lambda \}|,
\end{eqnarray}
where $| |$ is the set size operator.

The schematic diagram shown in Fig. \ref{fig_clusters} illustrates the cluster measurement algorithm described by Eqs. \ref{eq:phi_set} - \ref{eq:s_size}. In the provided example, there are four detected clusters labeled by $\lambda \in \{ 1,2,3,4 \}$. The set $\Phi_2$ contains the azimuthal positions of the $\lambda=2$ cluster at a chosen radial position $\rho'$, the set $R_2$ contains all the $\phi$ positions belonging to this cluster, $\delta\phi_2=3$ and $\delta\rho_2 = 16$ are respectively the largest thickness and the total radial extent of the cluster, and $s_2 = 38$ is the area of the cluster in pixels.

The analysis of a more detailed spatial structure of the interpolated features is further performed in rectangular coordinates $x_\lambda$ and $y_\lambda$ describing the Cartesian positions of the nodal points along each feature:

%\begin{equation}
%\label{eq:}
%F(I_k^{th}) = p_k
%\end{equation}

\begin{eqnarray}
\label{eq:rect_coord}
x_{\lambda, i} &=& \rho_{\lambda, i} \cos \phi_{\lambda, i}^* + x_0 \\
y_{\lambda, i} &=& \rho_{\lambda, i} \sin \phi_{\lambda, i}^* + y_0,   
\end{eqnarray}
in which $x_0$ and $y_0$ are the $x$ and $y$ position of the solar disk center obtained from the original image file and $i$ is the index of the nodal point as before. The obtained rectangular positions are used to compute the POS angle characterizing spatial orientation of each segment of the feature:

\begin{equation}
\label{eq:feature_angles}
\xi_{\lambda, i} = \tan^{-1}\left( \frac{y_{\lambda,i+1}-y_{\lambda,i}}{x_{\lambda,i+1}-x_{\lambda,i}} \right).
\end{equation}

The angle $\xi_{\lambda,i} \in [0, 2\pi]$ describes the local inclination of the $i^{th}$ segment of the feature with the label $\lambda$.  As confirmed in the following sections, these angles can be used as a proxy to the POS orientation of the coronal magnetic field between the two image locations, $x_{(\lambda,i},y_{\lambda,i})$ and $(x_{\lambda,i+1},y_{\lambda,i+1})$. 

To characterize the departure of the feature from the pure radial direction, we also compute the relative angles 

\begin{equation}
\label{eq:feature_rad_angles}
\xi_{\lambda, i}^{rad} = \tan^{-1}\left( \frac{y_{\lambda,i}-y_0}{x_{\lambda,i}-x_0} \right) \in [-90^\circ, 90^\circ].
\end{equation}

The relative angle equals zero if the feature is aligned with the radial direction at a given node. Positive (negative) $\xi_{\lambda, i}^{rad}$ value indicates that the feature segment is rotated counterclockwise (clockwise) compared to the radial direction. This angle is a convenient metric for selecting quasi-radial coronal features appearing in the open-field corona, and discarding those with large $|\xi_{\lambda, i}|$ which are more likely to belong to closed coronal loops. 

If the coronal image is studied in conjunction with a coronal magnetic field model, the discrepancy between the QRaFT features and the POS magnetic field is quantified by the local misalignment angle $\theta$ (Eq. \ref{eq:b_angles}). This angle is measured relative to the outward magnetic field interpolated at the location of the $i^{th}$ node. The outward field is constructed by multiplying the POS magnetic field $\vec{B}_{\lambda,i} = (B_{x}, B_{y})$ with the step function 

\begin{equation}
\label{eq:h}
h_{\lambda,i} = 2 \mathcal{H}(\vec{B}_{\lambda,i} \cdot \vec{r}_{\lambda,i}) - 2 = \left\{
    \begin{array}{ll}
        +1, \, \vec{B}_{\lambda,i} \,\, \mbox{is outward} \\
        -1, \, \vec{B}_{\lambda,i} \,\, \mbox{is inward,} \\
    \end{array}
\right.    
\end{equation}
where $\vec{r}_{\lambda,i} = (x_{\lambda,i} - x_0, y_{\lambda,i}-y_0)$ is the local POS radius vector. The misalignment angle is computed using the cross product of the outward magnetic field and the unit vector $\hat{f}_{\lambda, i} = (\cos \xi_{\lambda, i}, \sin \xi_{\lambda, i})$ specifying the local tangential direction along the feature:

\begin{equation}
\label{eq:b_angles}
    \theta_{\lambda, i} = \sin^{-1} \left( \frac{\hat{f}_{\lambda, i} \times (h_{\lambda,i} \, \vec{B}_{\lambda,i}) }{|\vec{B}_{\lambda,i}|}\right). 
\end{equation}

Since the feature nodes are indexed in the ascending $\rho$-order, the estimated $\theta_i$ values vary between $\pm 90^\circ$, with the positive misalignment angles signaling the departure of the feature from the magnetic field in the clockwise direction.

By combining the  angles $\xi$, $\xi^{rad}$ and $\theta$ evaluated at all the locations along all the features obtained with various combinations of the detection parameters $I_k^{th}$ and $\rho_j^{\min}$, we achieve an improved spatial coverage of the coronal regions containing identifiable structures. If a given image location contains more than one feature, the arithmetic average of the angles obtained using different detection settings is used in the output data arrays.

\begin{figure}
\begin{center}
\includegraphics[width=9 cm]{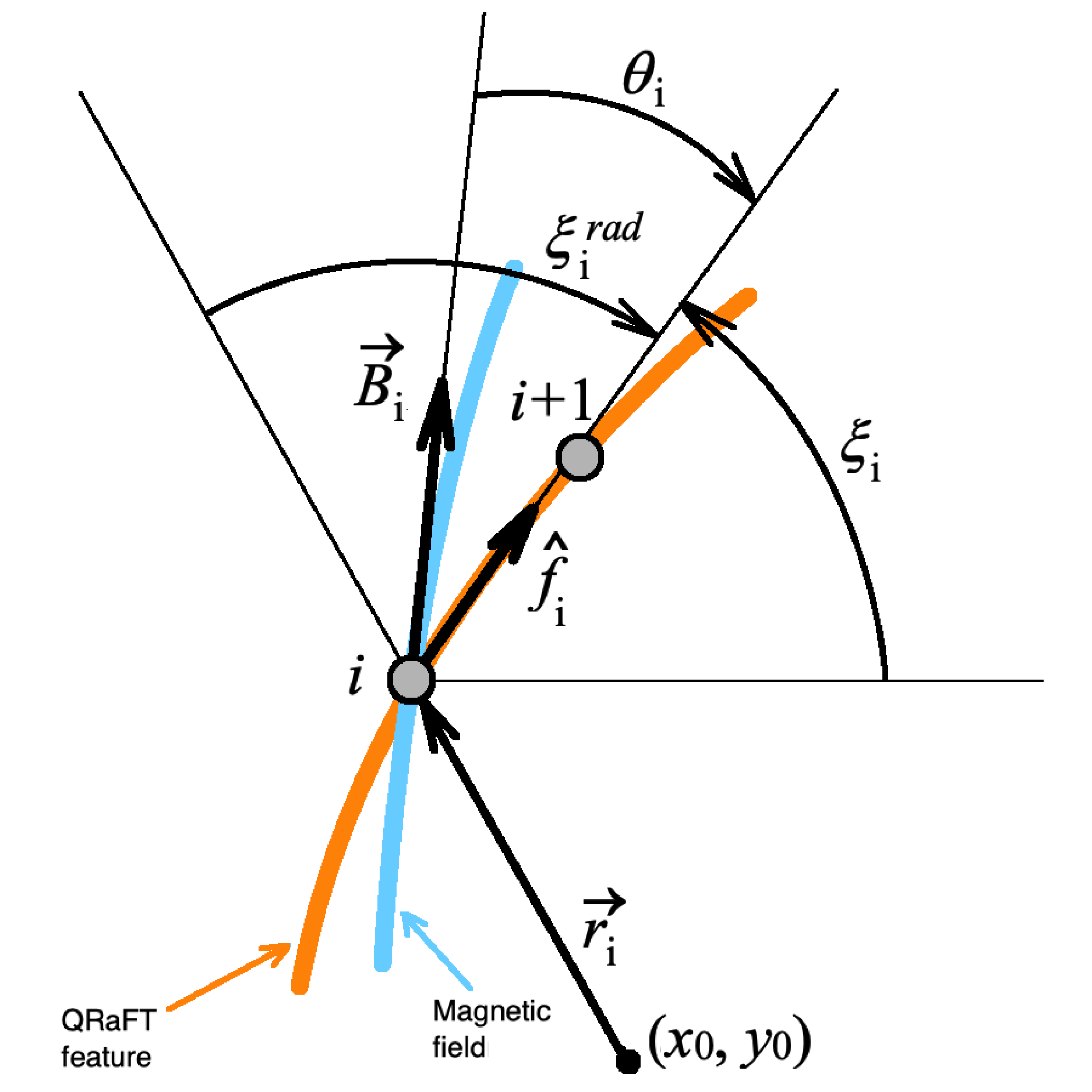}
\caption{\label{fig_angles} Schematic diagram showing the angular metrics (\ref{eq:feature_angles}) - (\ref{eq:b_angles}) at the location of a selected feature node. See text for details.}
\end{center}
\end{figure}

See Fig. \ref{fig_angles} for a graphical illustration of the angular metrics described in this section.

%-----------------------------------
%-----------------------------------

\subsection{Automatic filtering}
\label{sec:QRaFT_filtering}

The QRaFT feature metrics described above are used for an automatic filtering which allows us to eliminate unreliable or irrelevant detection results. The filtering includes several conditions which need to be fulfilled simultaneously for a feature to be classified as valid. 

The first two filtering conditions are applied to the thickness and the length of the features: 
\begin{eqnarray}
\label{eq:phi_cond}
\delta \phi_\lambda &\in& [\delta \phi_{\min}, \delta \phi_{\max} ] \\
\label{eq:rho_cond}
\delta \rho_\lambda &\in& [\delta \rho_{\min}, \delta \rho_{\max} ] 
\end{eqnarray}

Retaining the features with sufficiently small $\delta\phi$ is important because they communicate the most accurate information about the local magnetic filed orientation. However, if the azimuthal size is too small, the angular feature metrics become unreliable due to poor statistics. The upper limit on $\delta\rho$ is usually not as important -- in fact, the longest features often produce the most consistent tracing results, whereas applying the lower limit is critical because the shortest features do not allow for accurate determination of the orientation angles, and, in many cases, produce detection artifacts. 

The sufficient feature length is also ensured by requiring a large enough number of nodal points in the feature:

\begin{equation}
\label{eq:n_cond}
n_\lambda > n_{\min}.    
\end{equation}
 
To exclude spurious features characterized by low signal-to-noise ratios, we estimate the average brightness of the feature:

\begin{equation}
\label{eq:brightness}
I_\lambda = \frac{1}{n_\lambda+1} \sum_{i=0}^{n_\lambda} I(x_{\lambda,i}, y_{\lambda,i})    
\end{equation}

According to this definition, $I_\lambda$ is the average DN pixel count of the original image over the locations traced by the feature. To exclude the features described by the lowest optical signal, we applied a lower threshold to the $I_\lambda$ values:

\begin{equation}
\label{eq:I_cond}
I_\lambda > I_{\min}
\end{equation}

It is also important to focus the analysis on quasi-linear features which are characteristic of open-field regions. For this purpose, we calculated the root-mean-square value of the feature angles: 

\begin{equation}
\label{eq:sigma}
\sigma_\lambda = \sqrt{\langle \xi(x_{\lambda}, y_{\lambda})^2\rangle - \langle \xi(x_{\lambda}, y_{\lambda})\rangle^2}
\end{equation}

A near-zero $\sigma_\lambda$ is expected when the feature is reasonably approximated by a straight-line segment a high $\sigma$ indicates a large dispersion of the local angles indicating a departure from the linear shape. For filtering purposes, we used the angle spread per unit feature length, which can be seen as a rough proxy to the characteristic curvature of the feature:

\begin{equation}
\label{eq:curv_cond}
\frac{\sigma_\lambda}{n_\lambda} < c_{\max}
\end{equation}

The condition (\ref{eq:curv_cond}) ensures that the curvature of the feature is smaller than the upper curvature limit $c_{curv}$. It is applied to $\sigma_\lambda$ normalized by the number of nodes in the feature. This ratio is the largest for small features described by a large angle variability, which often represent detection artifacts or belong to closed coronal loops which are beyond the scope of QRaFT. 

\subsection{Code implementation}

The methods and algorithms described in the sections \ref{sec:QRaFT_enh} - \ref{sec:QRaFT_filtering} have been implemented in Interactive Data Language. The QRaFT package is available on the GitHub platform \citep{uritsky2024}. Table \ref{tab:subroutines} in the Appendix provides the complete list of the subroutines included in the package, while Table \ref{tab:keywords} lists the processing keywords enabling an in-depth customization of QRaFT runs, optimizing the performance  for a specific data source. Each subroutine has a detailed header comment with the information about the subroutine's purpose, arguments, return values, data formats, and application examples. The last three columns of Table \ref{tab:keywords} contain the values of the keywords used for processing each of the image types described below.

\section{Performance assessment}

\label{sec:performance}

To test the QRaFT methodology, three types of coronal images were used: (1) artificial white-light images produced by a global MHD model, (2) space-borne coronagraph images obtained from the STEREO COR1 instrument, and (3) ground-based coronal images obtained during a TSE. The following three subsections present selective processing examples for each type of image; a more systematic performance and error analysis of the methodology as part of a larger validation framework is reported in \citet{rura2025}. 

\subsection{Processing synthetic coronal images}
\label{sec:performance_MAS}

The synthetic coronal images described in this section were produced using a coronal plasma density distribution in a high-resolution MHD simulation, which enabled a direct quantitative comparison of the QRaFT features with the underlying magnetic geometry, since both the density and the magnetic field data were produced by the same numerical solution.

To construct the synthetic images, we used an output of the  Magnetohydrodynamic Algorithm outside a Sphere (MAS) model of the global solar corona \citep{mikic1999, lionello2009, mikic2018}. MAS implements a ``thermodynamic'' MHD approach with an energy equation that includes coronal heating, parallel thermal conduction, radiative loss, and Alfv\'en wave acceleration, making it possible to describe the interplay between magnetic and hydrodynamic forces, which is essential for modeling the production of the solar wind in the open-field corona. 

We used a MAS simulation constructed to predict the structure of the solar corona during the August 21, 2017 total solar eclipse. To achieve a realistic representation of the corona, the MAS simulation run was subject to a boundary condition derived from the observed photospheric magnetic field. The simulation employs a WTD approach to specify coronal heating \citep{lionello2014, downs2016}, and an energization technique for driving field-aligned currents over large-scale polarity inversion lines \citep{mikic2018, yeates2018}. The inner boundary condition was obtained from a handmade splice of synoptic map data from the HMI instrument \citep{scherrer2012} onboard the SDO spacecraft \citet{} over about 10 days prior to the eclipse. The boundary map included data for CR 2192 and near-real-time data from a part of the next rotation, CR 2193. To reproduce open-field plume-like structures at the poles, the polar caps were also filled with a random flux distribution whose net flux matches observations of the average net flux measured at high latitudes. The boundary magnetic field measurements span a time-range of approximately July 16 to August 11, 2017. 
The simulation is fully described by \citet{mikic2018}.

As mentioned earlier in the text, the key advantage of using an MHD simulation for testing feature-tracing codes, compared to solar imaging data, is the availability of ground-truth magnetic field data in the simulation. To test the performance of QRaFT, we constructed two types of synthetic images. The first type is 
a two-dimensional density array describing the distribution of the simulated plasma in a plane passing through the center of the Sun. The orientation of the central plane was chosen to be parallel to the POS of the STEREO COR1 telescope at the time when the image studied in the next section was collected. The central plane density is a highly artificial proxy to the white light emission from a single cross-section plane. Since the structure of this two-dimensional field is strongly constrained by the local magnetic field (see Section \ref{sec:QRaFT_general}), it can be used to investigate the inherent uncertainties of QRaFT tracing unaffected by more complex error sources such as projection and scattering effects. 

The second, more sophisticated type of synthetic coronagraph data studied here involves a LOS integration of the three-dimensional plasma density with the FORWARD code \citep{gibson2016} producing a realistic polarized brightness (pB) image for the chosen POS. Since the structure of the synthetic pB image is influenced by the LOS projection ambiguity and Thomson scattering geometry, its morphology is expected to deviate from the POS magnetic field direction, and the degree of this distortion can be quantified by QRaFT. 

The morphology of the synthetic images was investigated by QRaFT, which detected an extended ensemble of quasi-radial optical features around the Sun.  The local orientation angles describing these features were compared with the directions of the magnetic field lines in the same MAS run. The misalignment angles representing the discrepancy between the QRaFT features and the MAS magnetic field provided a basis for a robust assessment for the accuracy of the image-processing method. 

%-----------------------------------
\begin{figure*}%[h]
\begin{center}

\includegraphics[width=8.6 cm]{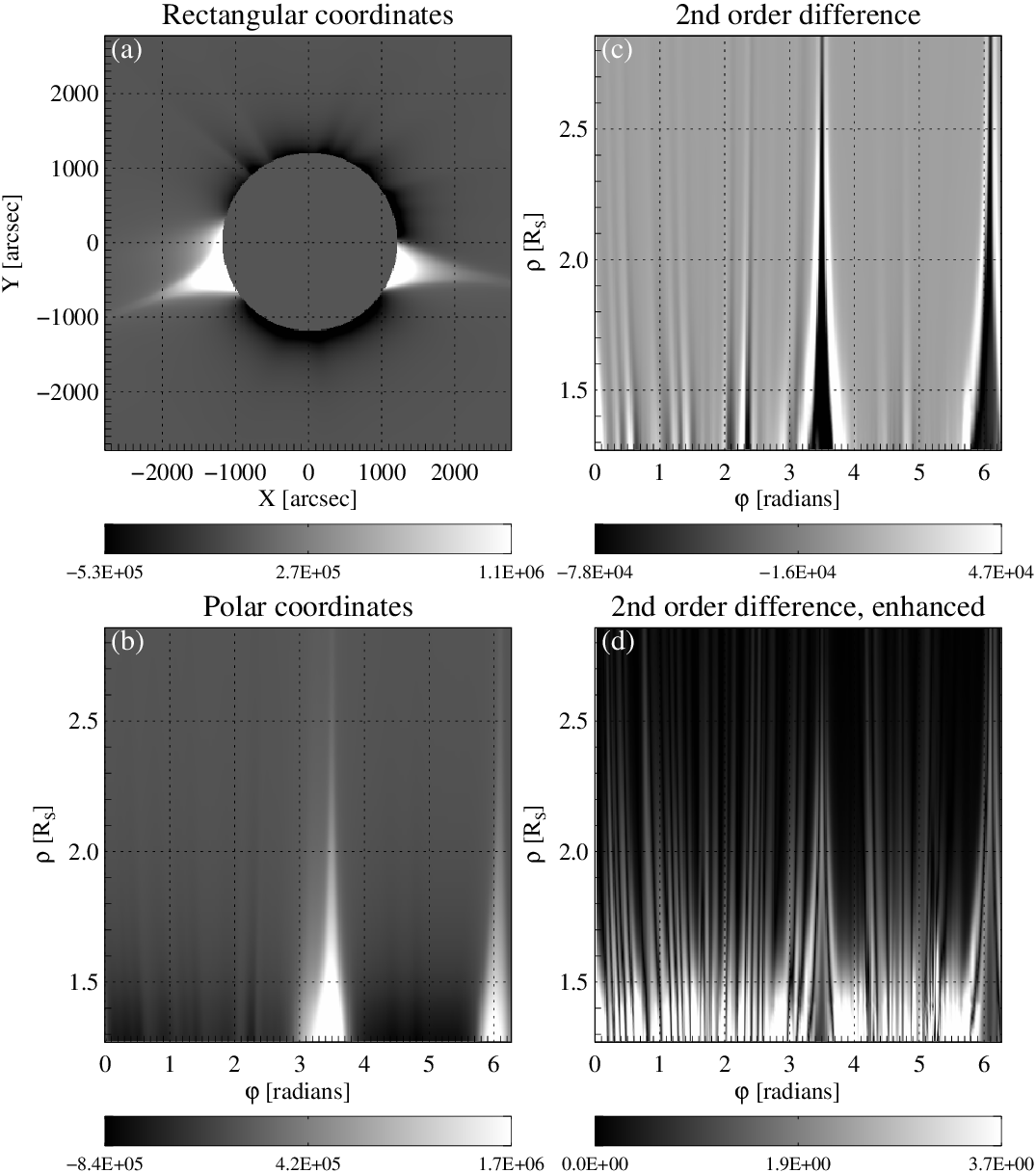} \,\,\,\, %\includegraphics[width=8.7 cm]{fig_images_pB_.eps}
\includegraphics[width=9.0 cm]{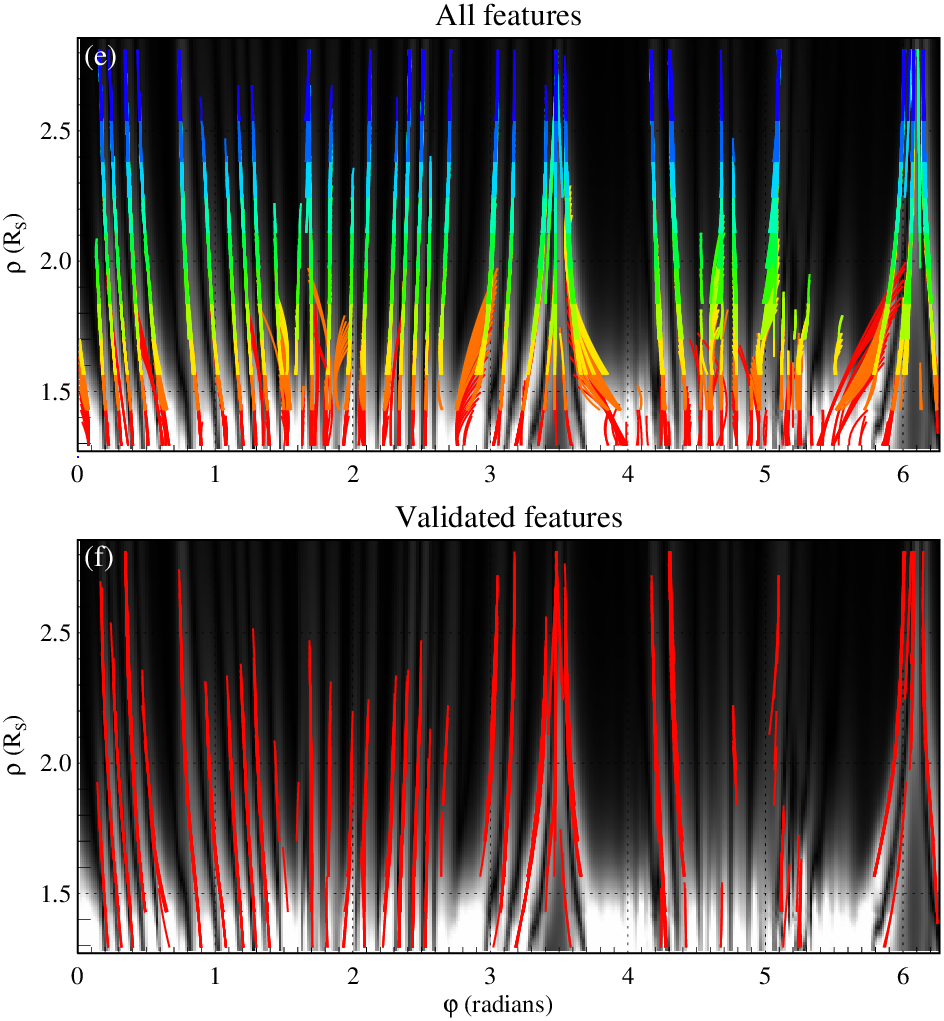}
\caption{\label{fig_MAS_ne_images} QRaFT segmentation of the central-plane density array in the MAS model. (a) Processed image array $I_{detr}$ (\ref{eq:detr}) in Cartesian coordinates after radial detrending and smoothing; (b) transformed array $I(\phi,\rho)$ (\ref{eq:polar}) in plane polar coordinates; (c) the signed second-order azimuthal difference array $I_{diff}$ (\ref{eq:diff}); (d) the unsigned second order difference array $I_{enh}$ (\ref{eq:enh}) after azimuthal detrending; (e) Tracing unsigned second-order azimuthal gradients, with the colors marking ten tracing runs with different inner boundary $\rho^{min}$; (f) final tracing product obtained by combining the individual tracing runs and automatically removing unreliable features as explained in Section \ref{sec:QRaFT_filtering}. }
\end{center}
\end{figure*}
%-----------------------------------

Fig. \ref{fig_MAS_ne_images} presents an example of a QRaFT-based analysis of a central-plane plasma density obtained from the studied MAS model run. Fig. \ref{fig_MAS_ne_images}a shows the pre-processed density array $I_{detr}$ in Cartesian coordinates. The array was subject to the radial detrending procedure (\ref{eq:detr}) reducing the average decay of the density with the heliocentric distance and making the density structures geometrically more consistent and easier to detect. Fig. \ref{fig_MAS_ne_images}b shows the detrended density image transformed into plane polar coordinates (Eq. \ref{eq:polar}). It can be seen that in either coordinate system, the detrended image is relatively smooth and it does not contain much of a small-scale azimuthal structure. This structure becomes much more apparent after the azimuthal finite differencing (\ref{eq:diff}) and detrending (\ref{eq:enh}). The results of these processing steps are shown in Fig. \ref{fig_MAS_ne_images}c and \ref{fig_MAS_ne_images}d, correspondingly. 

The fine quasi-radial structure of the MAS density array visualized in Fig. \ref{fig_MAS_ne_images}d is expected to be aligned with the magnetic field (see Section \ref{sec:QRaFT_general}), and it has been used for the tracing purposes. Fig. \ref{fig_MAS_ne_images}e shows 10 sets of tracing results obtained using 10 different positions of the inner boundary $\rho^{min}$. The traced lines represent the polynomial interpolation (\ref{eq:phi_poly}) to the polar coordinates (\ref{eq:rho_array})-(\ref{eq:phi_array}) of the connected pixels clusters detected using the adaptive thresholding methodology (\ref{eq:lbl}). Not all the features  plotted in Fig. \ref{fig_MAS_ne_images}e have a shape consistent with the quasi-radial structures of interest -- some of the plotted lines have an excessive curvature, spurious location, and/or a tilt angle which seem to contradict the local geometry of the image. These artifacts have been successfully removed by applying the validation filters (\ref{eq:phi_cond}) -(\ref{eq:curv_cond}). The final, validated set of traced structures is shown in Fig.\ref{fig_MAS_ne_images}d and it can be seen that it is in reasonable agreement with the underlying image $I_{enh}$. 

%-----------------------------------
\begin{figure*}%[h]
\begin{center}

\includegraphics[width=8.6 cm]{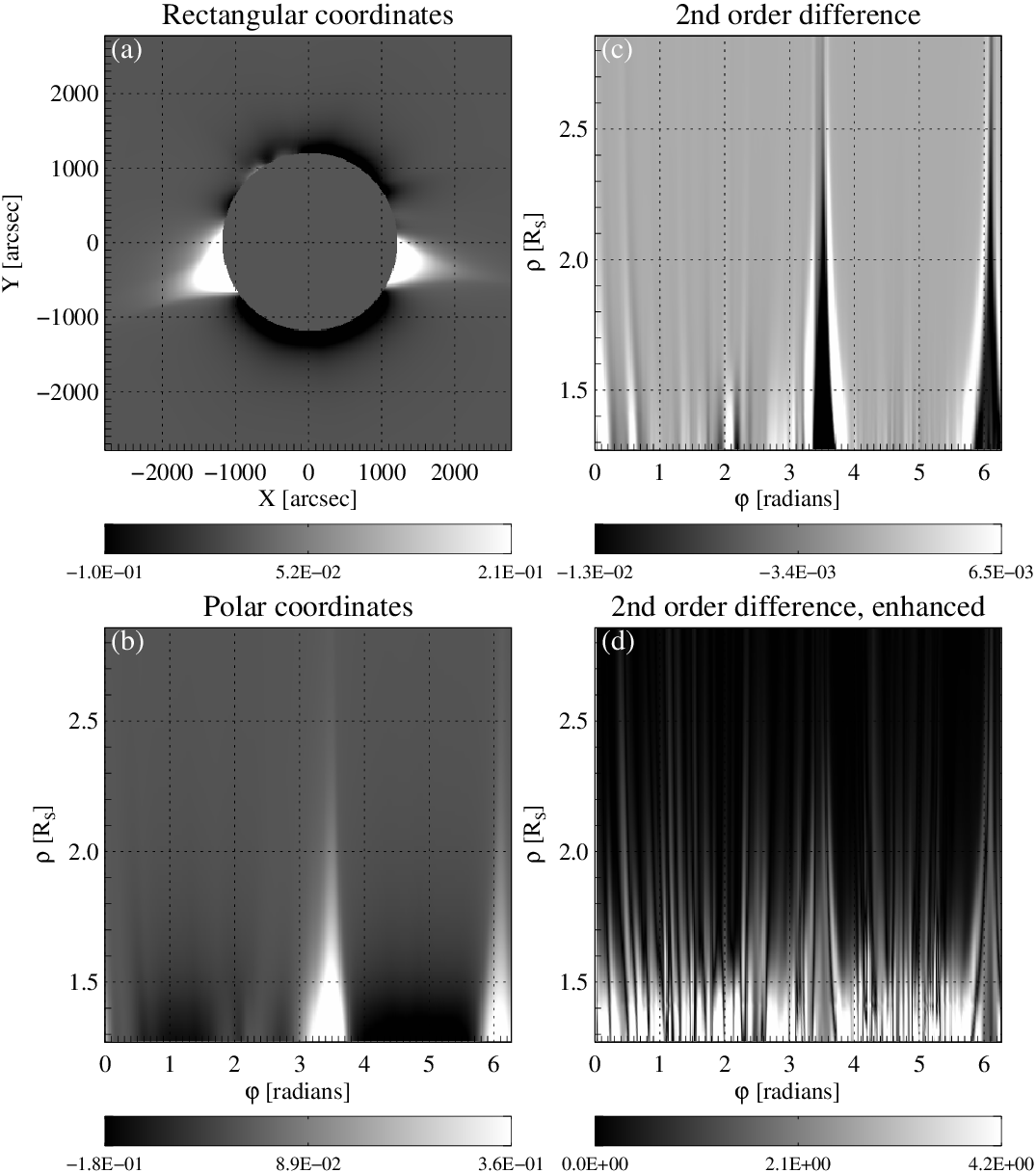} \,\,\,\, \includegraphics[width=9.0 cm]{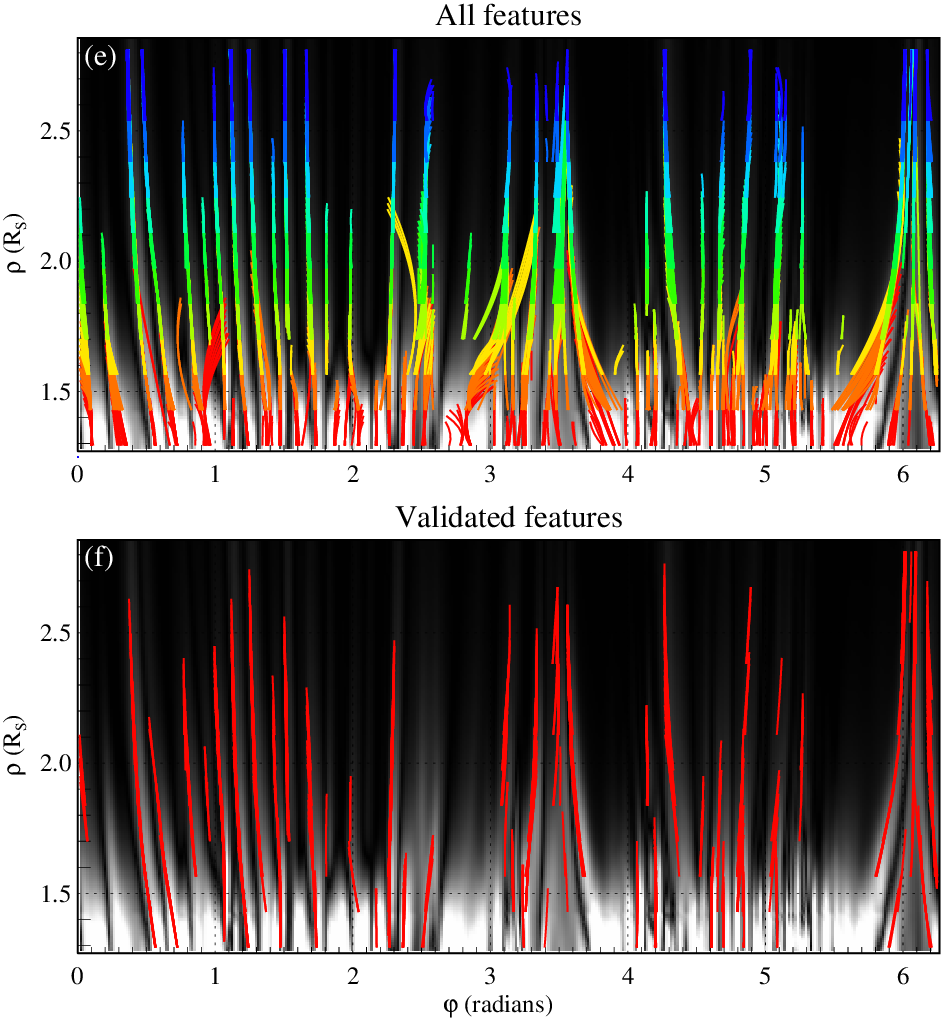}
\caption{\label{fig_MAS_pB_images} QRaFT segmentation of a synthetic pB image produce by MAS and FORWARD codes. See Fig. \ref{fig_MAS_ne_images} for notations. }
\end{center}
\end{figure*}
%-----------------------------------

Fig. \ref{fig_MAS_pB_images} shows an example of QRaFT processing of a synthetic pB image obtained from the same MAS run and the same virtual vintage point as the central-plane density image shown in Fig. \ref{fig_MAS_ne_images}. The information is also presented in the same format. As with the density image, the enhanced pB image reveals an abundant quasi-radial structure. The global patterns of both structures are similar, which suggests that the LOS projection effects and Thomson scattering distortions have limited impact on large coronal scales close to the Sun, at least for this particular view of the corona during the declining phase of the solar cycle. The small-scale difference is more significant, and it leads to a noticeably larger fraction of incorrectly traced features in the case of the synthetic pB image (Fig. \ref{fig_MAS_pB_images}e) compared to the density image (Fig. \ref{fig_MAS_ne_images}e). The pB features remaining after the validation (Fig. \ref{fig_MAS_pB_images}f) are well aligned with the studied image but they tend to be shorter and more sparse compared to the density features, reflecting the fact that a smaller percentage of pB features passes the filtering conditions (\ref{eq:phi_cond}) -(\ref{eq:curv_cond}).

\begin{figure*}%[h]
\begin{center}

\hspace{1.4cm} Plasma density \hspace{5.8cm} Polarized brightness
\includegraphics[width=8.0 cm]{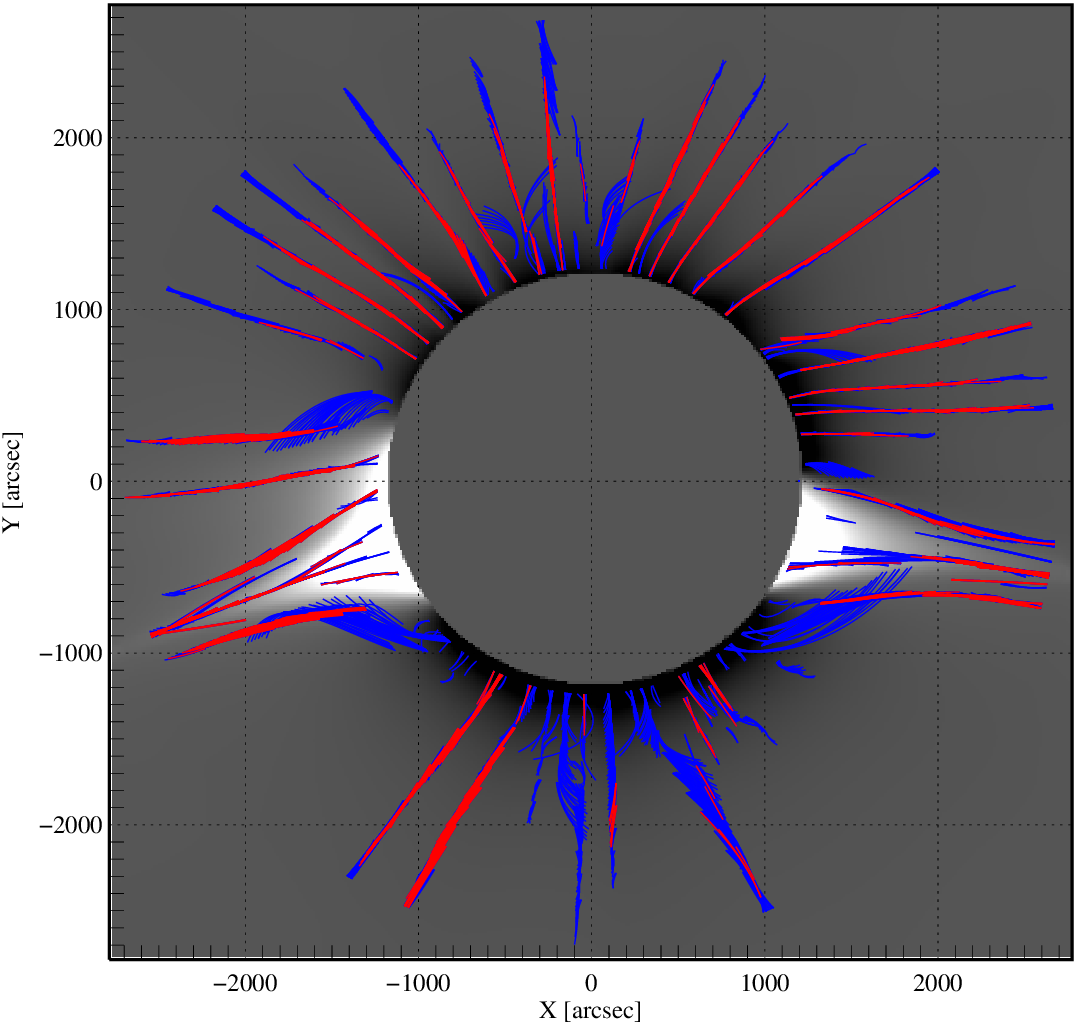} \,\,\,\, \includegraphics[width=8.0 cm]{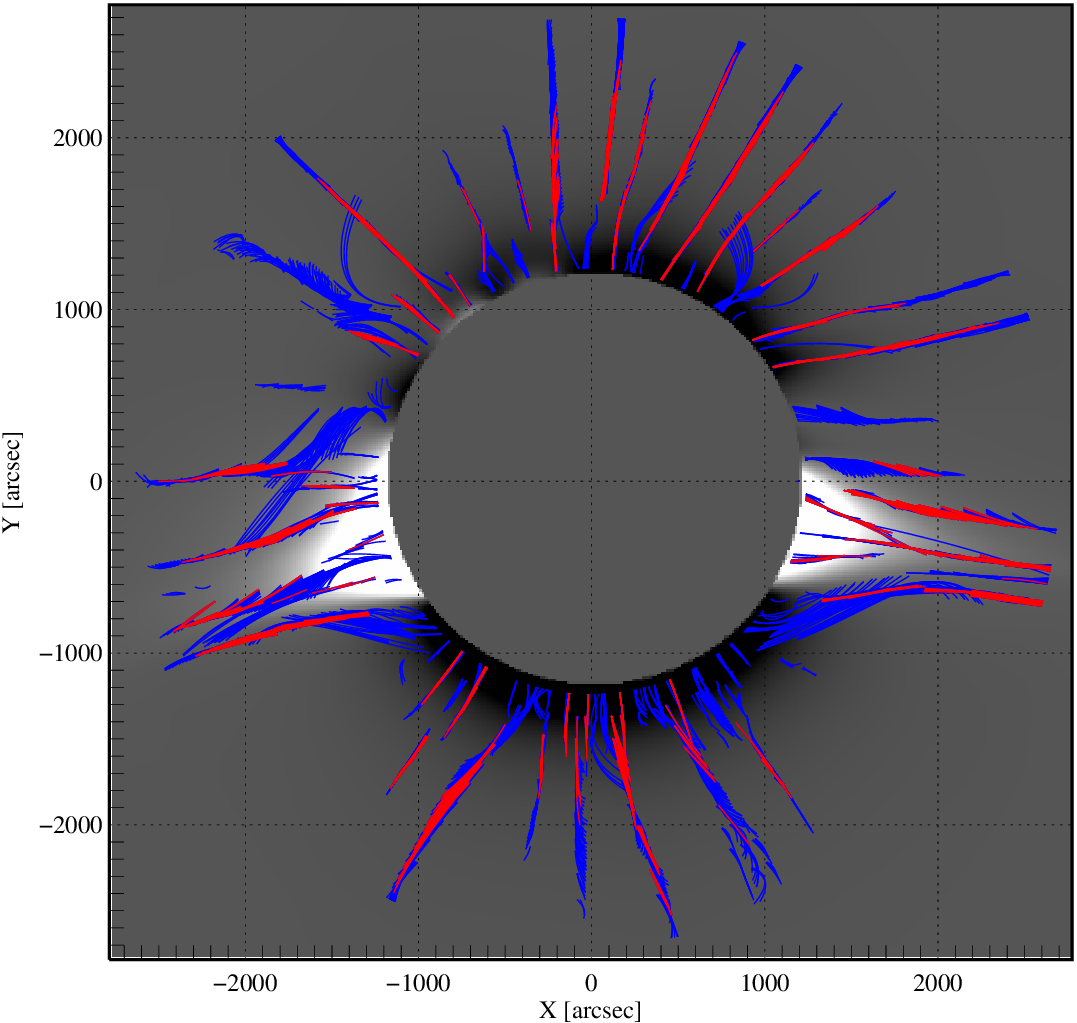}
\includegraphics[width=8.0 cm]{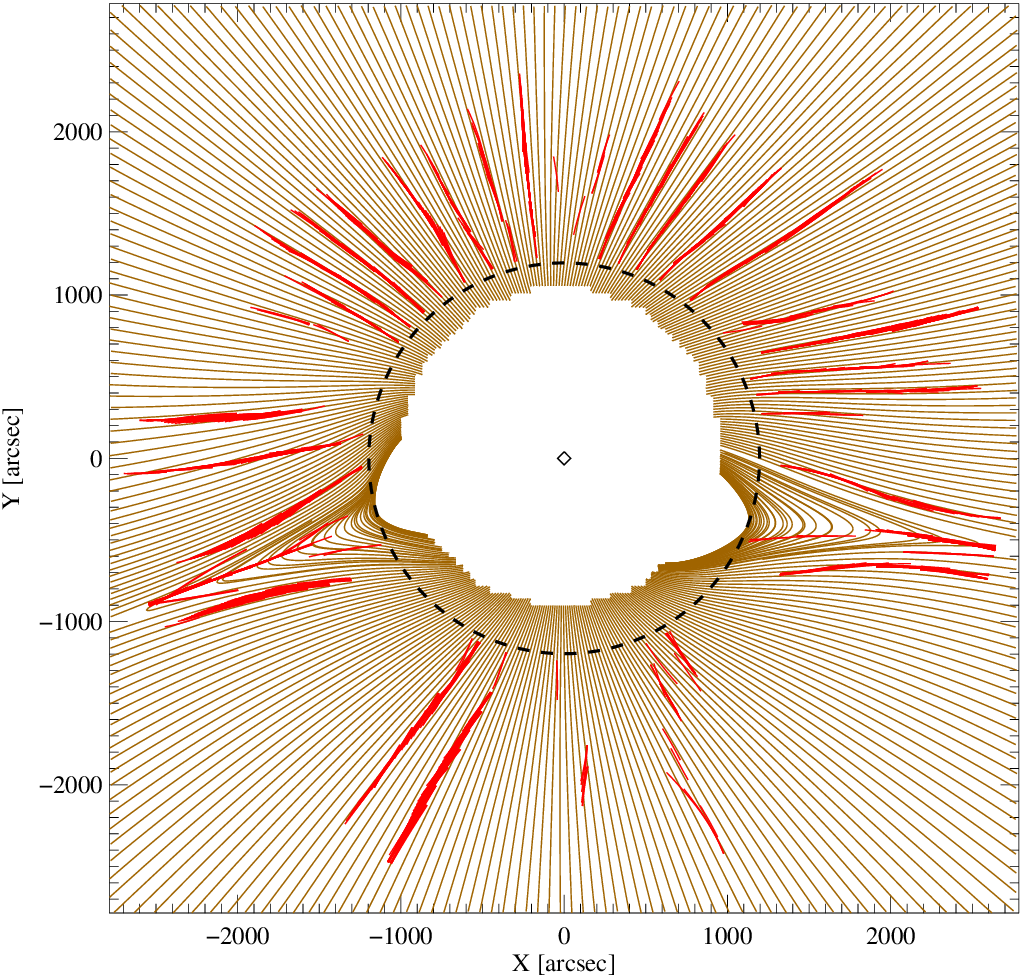} \,\,\,\, \includegraphics[width=8.0 cm]{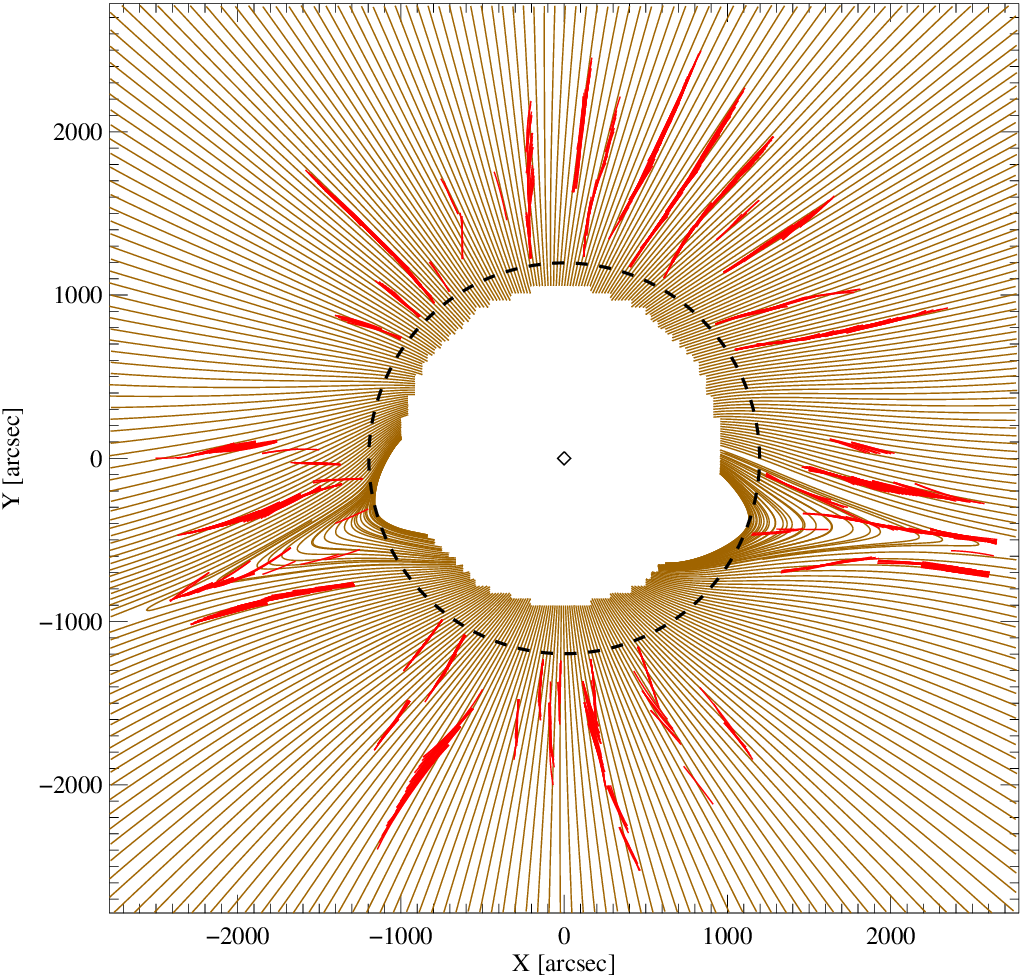}
\caption{\label{fig_MAS_alignment} Top panels: 
Automatic validation of the QRaFT features detected in the central plane density (left) and pB brightness (right) outputs of the MAS model. Red lines show the validated features that meet the filtering conditions (\ref{eq:phi_cond}) - (\ref{eq:curv_cond}) and are included in the final data product; blue lines show the discarded features excluded from the final QRaFT product. The processed enhanced images are shown on the background. 
Bottom panels: validated QRaFT features traced in the density array (left) and the synthetic pB image (right) obtained from the MAS model overplotted with the POS magnetic field lines (brown) in the same model.  }
\end{center}
\end{figure*}
%-----------------------------------

Fig. \ref{fig_MAS_alignment} provides a deeper insight into the process of QRaFT segmentation and validation. The first row of panels shows the QRaFT features reprojected into the rectangular coordinates and overplotted with the detrended synthetic images discussed above. Red lines show the validated features retained after the filtering step; blue lines are the spurious features which did not pass the filtering conditions. The gray solid circle in the middle of the images is the region not used for the segmentation ($\rho < 1.33$ $R_S$ in these tracing runs). The bottom panels show the validated QRaFT features (red) superposed with the central-plane magnetic field lines of the MAS model (brown). Since both the synthetic images and the magnetic field represent the same state of the same  system, the two sets of lines are expected to be very closely aligned. It can be seen that this is indeed the case for most of the studied locations around the Sun, but some of the traced features intersect the magnetic field lines at significant angles. This tends to occur where the magnetic flux is closed rather than open, such in large-scale streamers. The presence of closed flux presents a challenge for the QRaFT algorithm since it was optimized for segmenting large-scale open magnetic geometries based on the azimuthal differencing. In a way, this problem is ``hardwired'' into the method, and it is the price that we pay for being able to segment quasi-radial coronal structures in very noisy images, as discussed in the next section. While this limitation is not fundamental and could be removed in the  future, it should be taken into account when working with the current release of the QRaFT package. 

%In general, it is not recommended to trust the tracing results inside or in an immediate proximity of closed-field coronal regions. %, such as e.g. closed loop systems of ARs and streamers (??).

\begin{figure*}
\begin{center}

%\hspace{0.4cm}  Plasma density
\includegraphics[width=14.0 cm]
{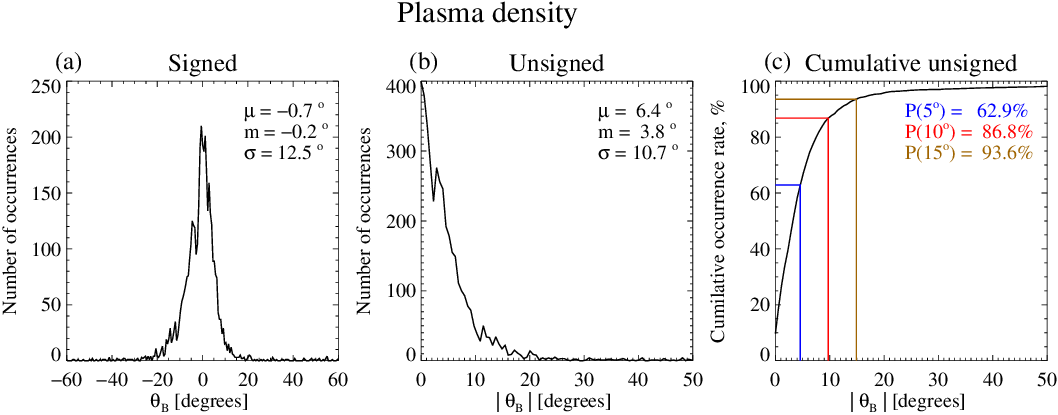} 

%\hspace{0.4cm}  Polarized brightness
\vspace{0.6cm}
\includegraphics[width=14.0 cm]{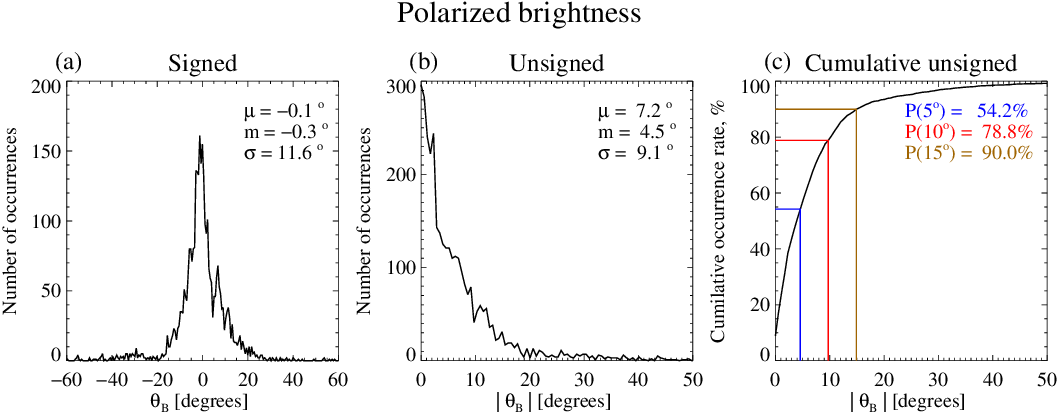}

\caption{\label{fig_MAS_stat} Misalignment angle statistics describing the discrepancy between the QRaFT features detected in the central-plane density array (top) and in the synthetic pB image (bottom) obtained from the MAS model, as compared to the POS orientation of the magnetic field in the same model. (a) The occurrence rate histogram of the signed angle misalignment $\theta_B$ angle; (b) the occurrence rate histogram of the unsigned angle $|\theta_B|$; (c) normalized cumulative distribution on the unsigned angle, showing occurrence rates for three  characteristic percentile levels. The parameters $\mu$, $m$ and $\sigma$ in panels (a-b) are respectively the sample mean, the median and the standard deviation of the corresponding sample.}
\end{center}
\end{figure*}
%-----------------------------------

Processing synthetic coronal images such as those produced by the MAS code provides a valuable opportunity to quantify the accuracy of the segmentation results using the angular metrics described in section \ref{sec:QRaFT_metrics}. Fig. \ref{fig_MAS_stat} shows the statistics of the misalignment angles $\theta$ (\ref{eq:b_angles}) in the QRaFT outputs. The angles characterize the discrepancy in the local orientation of a QRaFT feature relative to the model's magnetic field passing through a given interpolation node (Fig. \ref{fig_angles}). The first raw of the panels in Fig. \ref{fig_MAS_stat} presents several types of $\theta$ distributions of the central plane density image shown in Fig. \ref{fig_MAS_ne_images}; the second row is for the MAS-generated pB image in Fig. \ref{fig_MAS_pB_images}. 

By definition, $\theta > 0$ ($\theta < 0$) if the feature is rotated clockwise (counterclockwise) with respect to the $\vec{B}$ field line (disregarding the sign of the vertical magnetic field component), and it is close to $0^\circ$ if the feature is perfectly aligned with the magnetic field. The occurrence histograms of the signed misalignment angles shown in the left panels of Fig. \ref{fig_MAS_stat} exhibit sharp peaks at $\theta \approx 0$, indicating that most of the detected features are well aligned with the $\vec{B}$ field. The mean ($\mu$) and the median ($m$) values of the signed angles are quite small for both images. The standard deviation $\sigma$ of about $12-13^\circ$ suggests that there is a substantial spread in the performance of the tracing algorithm depending on the local coronal morphology. The approximately symmetric form of the occurrence distributions indicates the absence of a significant azimuthal bias in the tracing accuracy. 

The second panels in Fig. \ref{fig_MAS_stat} show the occurrence statistics of the unsigned misalignment angle $|\theta|$, which reveals additional details about the measurement uncertainty characterizing the QRaFT code. As in the case with the signed $\theta$ distributions, the unsigned distributions show a narrow peak near $\theta =  0$ indicating that the majority of the detected features a field-aligned. The distributions exhibit heavy tails, implying that the decay of the occurrence rate with $|\theta|$ is slower than exponential. Such statistical behavior is often a hallmark of a biased measurement error caused by a non-random factor, which could be an encounter of the QRaFT algorithm with closed magnetic loops resulting in disproportionally large tracing errors (see the discussion of Fig. \ref{fig_MAS_alignment} above).

The rightmost panels in Fig. \ref{fig_MAS_stat} showing the normalized cumulative distributions $P(\theta)$ of the unsigned misalignment angle characterize the statistical significance of the systematic tracing errors. The plots also show percentage occurrence rates for 3  error levels ($|\theta|$ = $5^\circ$, $10^\circ$ and $15^\circ$). These rates represent a relative fraction of the feature nodes characterized by a magnetic field misalignment angle smaller than the indicated error. The provided values show, for example, about 87\% of the locations of the plasma density features are field-aligned within the $10^\circ$, while only 78\% of the  pB tracing results are characterized by the $\theta$ angles lying below this error benchmark. Overall, the QRaFT features obtained from the synthetic pB image  demonstrate a systematically smaller fraction of accurately traced nodes compared to the central-plane density image, which is expected due to the geometric ambiguities caused by Thomson scattering \citep{gibson2016}. 

\subsection{{STEREO COR1} example}
\label{sec:performance_COR1}

To demonstrate the performance of QRaFT on a real-life coronal image from a space-borne coronagraph, we used a set of images from the Solar Terrestrial Relations Observatory-B (STEREO)/COR1 coronagraph \citep{kaiser2008, howard2008} collected on 2017-08-25. The image represents median values of the COR1 pixels in 50 COR1 pB images covering about 8 hours. Using a 50-frame median allowed us to reduce the pixel noise without significant losses of the sharpness of the quasi-static coronal structures targeted by QRaFT. We verified that the additional tracing uncertainty due to the solar rotation over the 8-hour interval is by at least one order of magnitude smaller than the typical QRaFT uncertainty, and this effect can therefore be neglected. After that, the image was spatially aligned with the model grid, which enabled a comparison of the observed features and the simulated magnetic field geometry. The mean vantage point of the COR1 image matches the MAS projection used to create the synthetic arrays described in Section \ref{sec:performance_MAS}. It has also been rebinned to match the resolution of the synthetic data, to enable direct comparison of the COR1 features with the MAS magnetic field.

%-----------------------------------
\begin{figure*}
\begin{center}

\includegraphics[width=8.6 cm]{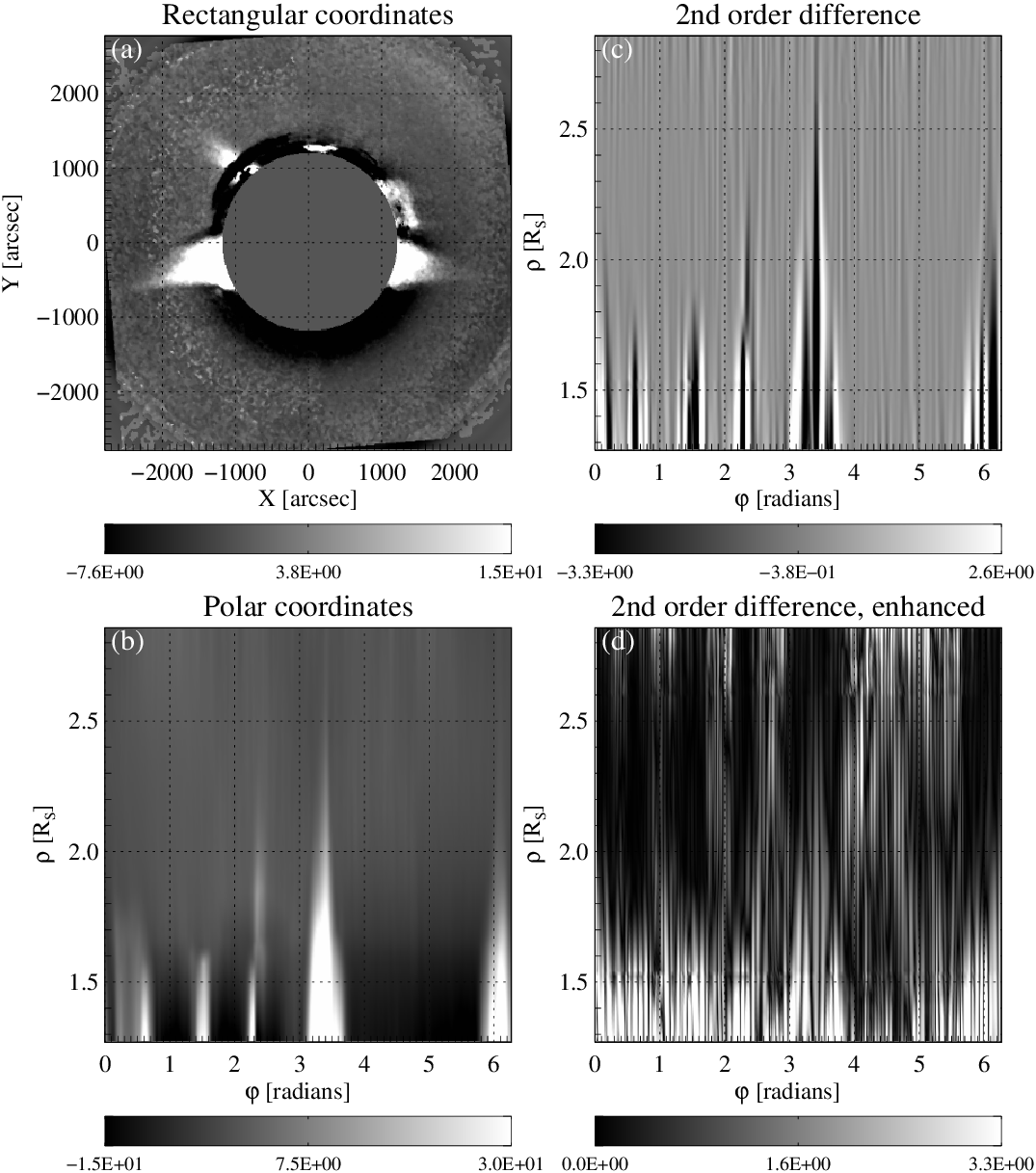} \,\,\,\, \includegraphics[width=9.0 cm]{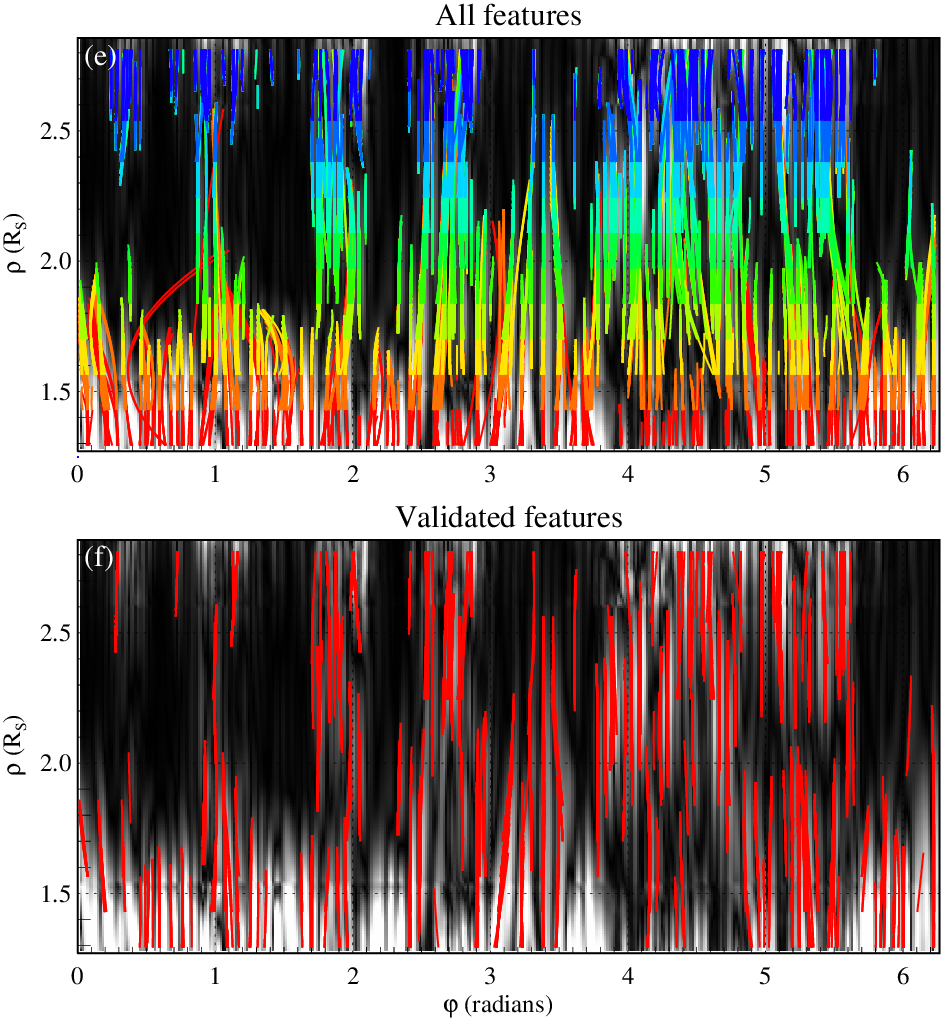}
\caption{\label{fig_COR1_images}  QRaFT segmentation of a COR1 pB image aligned with the synthetic MAS images shown in Section \ref{sec:performance_MAS}. See Fig. \ref{fig_MAS_ne_images} for notations.}
\end{center}
\end{figure*}
%-----------------------------------

Fig. \ref{fig_COR1_images} shows the processing stages of the COR1 image using the same format as in Figs. \ref{fig_MAS_ne_images} and \ref{fig_MAS_pB_images}. As expected, the real-life coronal image has a substantial level of pixel noise as well as some geometric artifacts which limit the accuracy of the radial and azimuthal detrending. These artifacts manifest themselves in the quasi-periodic large-scale gradients seen in the final enhanced image $I_{enh}$, in addition to the small-scale structure targeted by the tracing code. The distorted image enhancement led to a less accurate feature tracing: the set of the filtered QRaFT features in the COR1 image (Fig. \ref{fig_COR1_images}f) is a small fraction of all the detected features (Fig. \ref{fig_COR1_images}e). It can also be noticed that the validated features tend to be shorter compared to the features detected in the synthetic images. Overall,  QRaFT exhibits an arguably lower performance when processing COR1 images compared to the synthetic data, even though the MAS model is unable to reproduce the small-scaler coronal structure.  A higher-quality TSE coronal image discussed in the next section produced much better tracing results. In this context, the COR1 example presented here emphasizes the future need for more advanced space-borne coronagraph  products allowing for reliable reconstruction of the underlying magnetic geometry, which is a critical component of accurate space weather forecasting. 

%-----------------------------------
\begin{figure*}
\begin{center}
\includegraphics[width=15.0 cm]{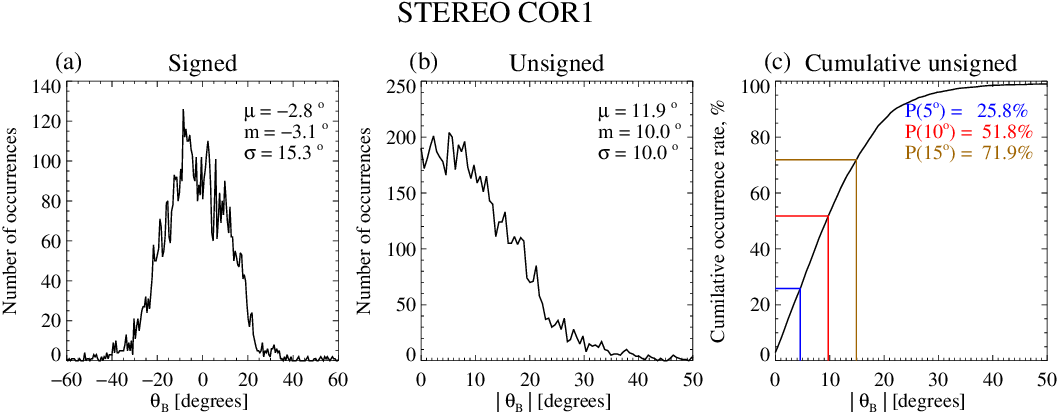}
\caption{\label{fig_COR1_stat} Misalignemnt angle statistics for the QRaFT features detected in the COR1 image. Notations are the same as in  Fig. \ref{fig_MAS_stat}.}
\end{center}
\end{figure*}

%-----------------------------------
\begin{figure*}[h]
\begin{center}

\includegraphics[width=8.6 cm]{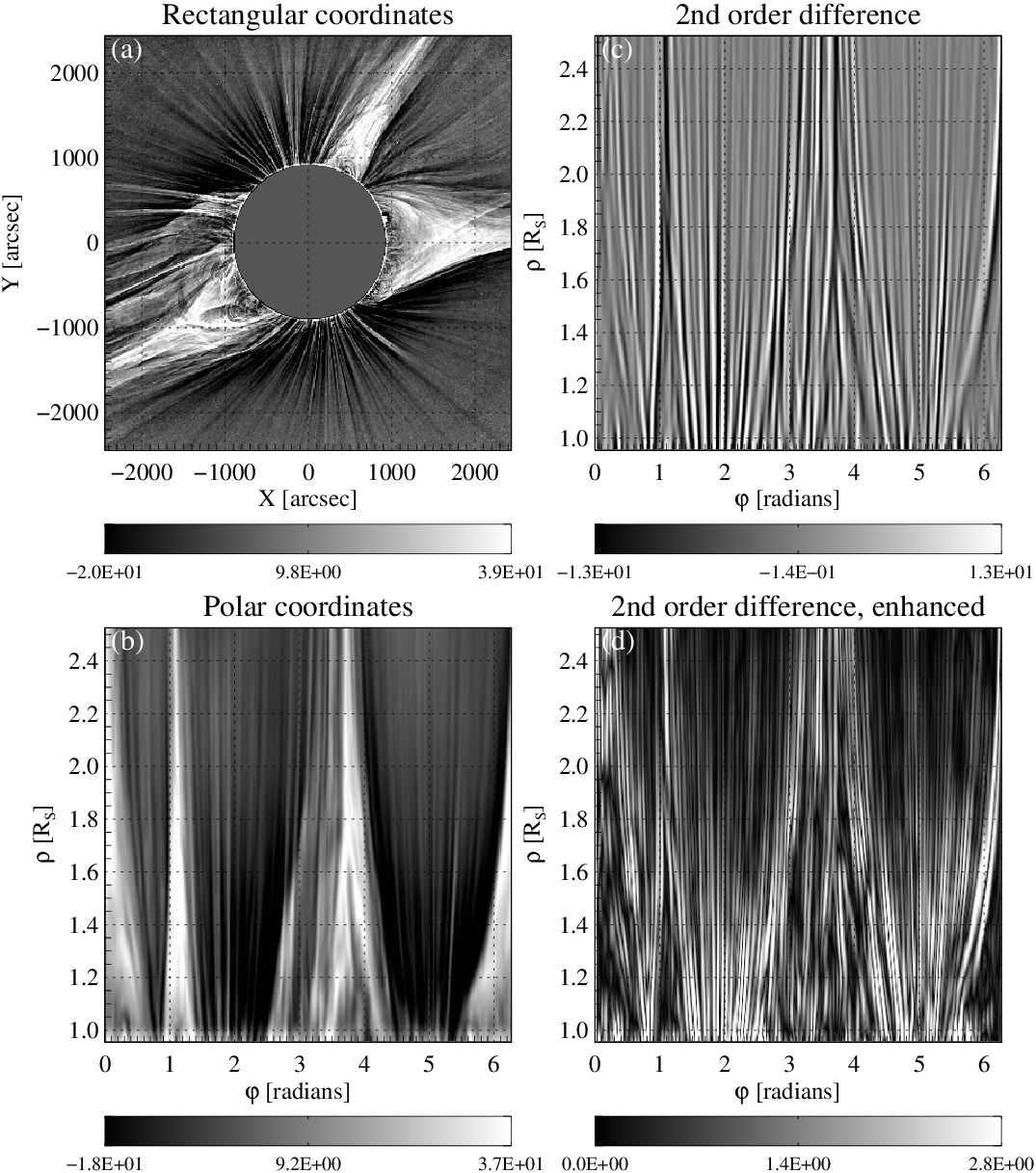} \,\,\,\, \includegraphics[width=9.0 cm]{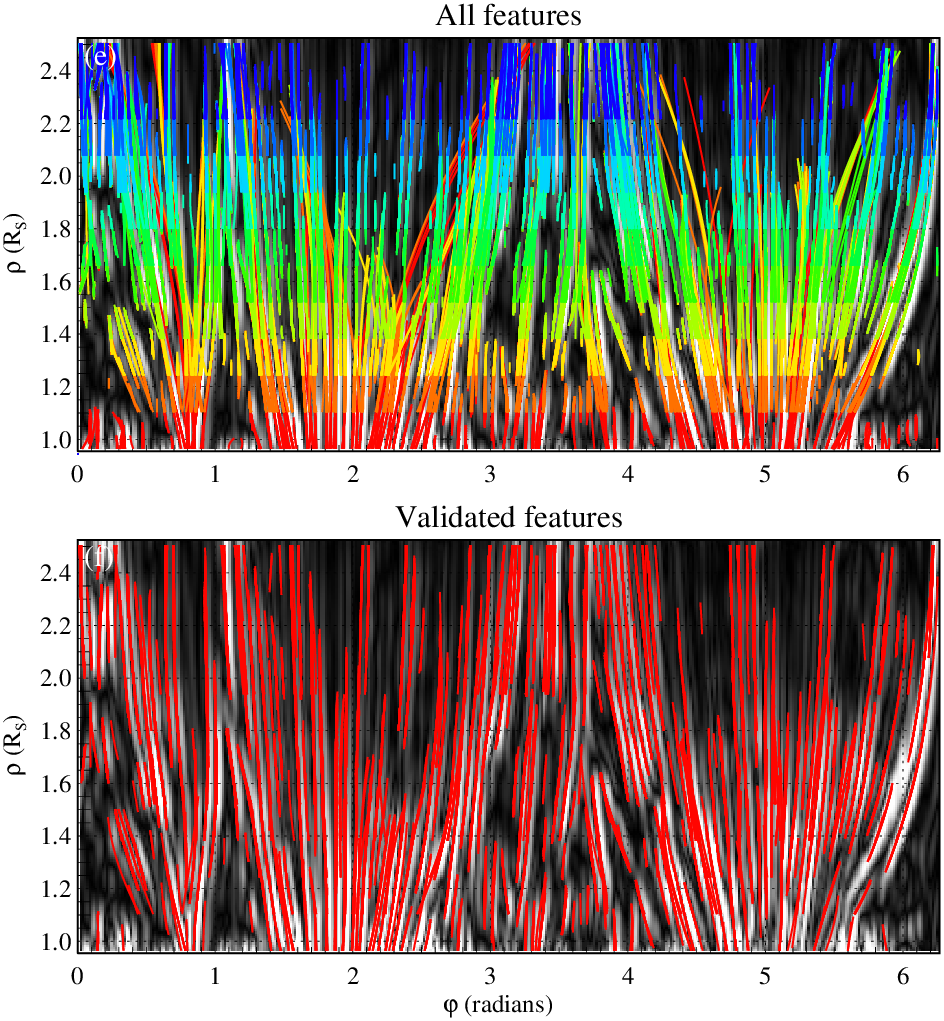}
\caption{\label{fig_TSE_images} QRaFT segmentation of a ground-based TSE image  (courtesy of M. Druckm{\"u}ller, P. Aniol and S. Habbal) acquired on 2017-08-21. See Fig. \ref{fig_MAS_ne_images} for notations.}
\end{center}
\end{figure*}
%-----------------------------------

\begin{figure*}[h]
\begin{center}
\includegraphics[width=10.0 cm]{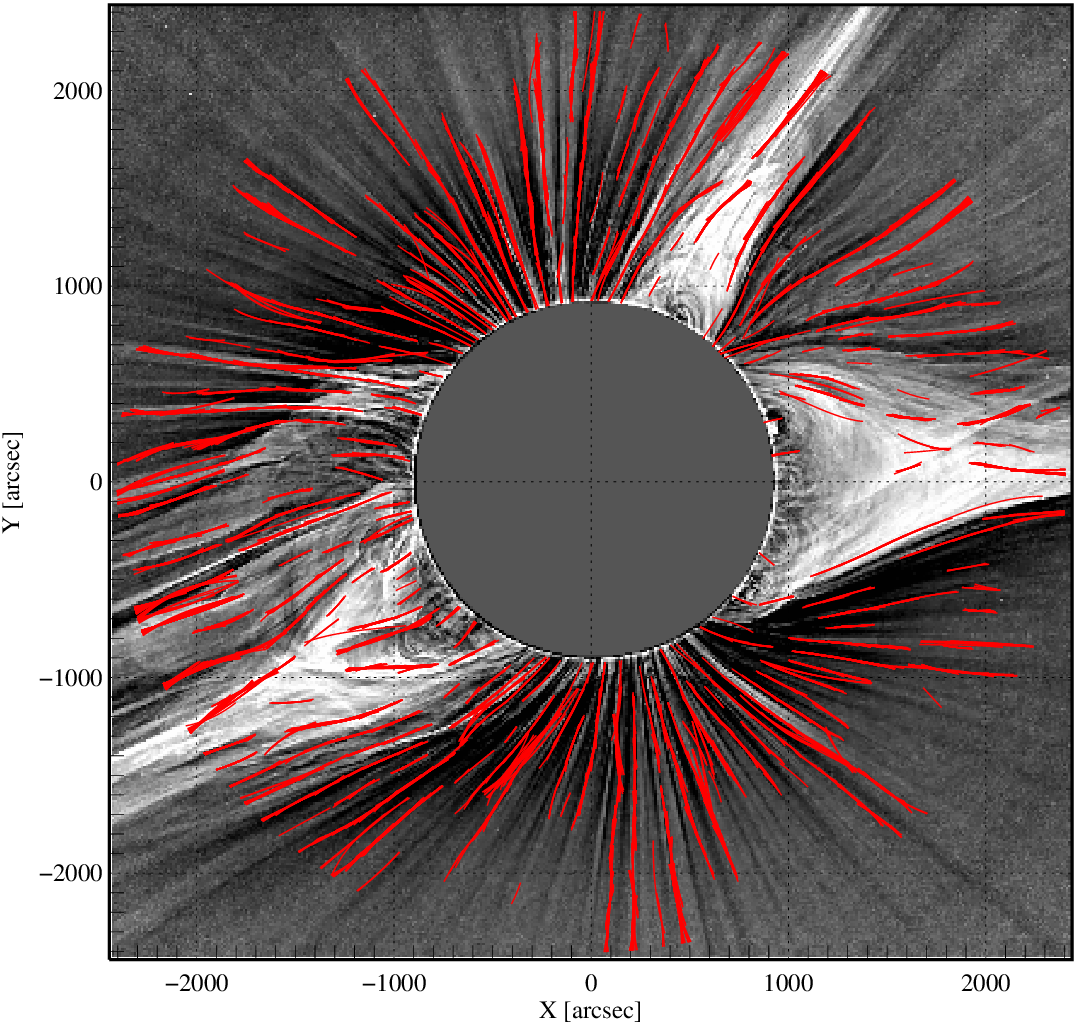}
\caption{\label{fig_TSE} Validated QRaFT features of the studied 2017-08-21 TSE image reprojected to the rectangular coordinate system. The background TSE image is courtesy of M. Druckm{\"u}ller, P. Aniol and S. Habbal. }
\end{center}
\end{figure*}
%-----------------------------------

Since the COR1 image was co-aligned with the model arrays, it was technically possible it compute the magnetic misalignment angles characterizing the COR1 image features the same way it was done for the QRaFT features in the synthetic images generated by the model. The histograms of the COR1  $\theta$ angles shown in Fig. \ref{fig_COR1_stat} confirm that the uncertainty of the orientation of the traced features relative to the MAS magnetic field is significantly higher than the uncertainty characterizing the synthetic data. The signed and unsigned $\theta$ histograms of the STEREO COR1 features are characterized by higher sample mean and median values as well as by substantially lower rate of accurately traced locations, with only 52\% of the feature nodes described by $|\theta|<10^\circ$. It should be noted, however, that the dominant cause of the spread of the misalignment angles in the COR1 data is most likely not the poor image quality but the large-scale inconsistency between the model and the real corona for this particular perspective. This may not be surprising given that the magnetic data used to prepare the boundary conditions of the model is at least 14 days old by this point. But if this is indeed the case, the quantitative metrics provided in Fig. \ref{fig_COR1_stat} should be interpreted as data-derived measures of the accuracy of the model run setup rather than performance benchmarks of the segmentation code. It is also worth mentioning that we are comparing here the POS magnetic field to features measured in LOS integrated observations, which is another possible source of discrepancy. Future investigations will explore using the emissivity-weighted average magnetic field components in the image plane.

\subsection{TSE image example}
\label{sec:performance_TSE}

TSE events provide a rare and unique opportunity of high-resolution imaging of coronal structures from the ground. With the Moon's shadow fully blocking the bright solar photosphere, the TSE conditions create a perfect natural ``coronagraph''. The TSE images tend to have high SNR and contrast resolution and contain an unprecedented amount of details across a large range of distances from the Sun.

Fig. \ref{fig_TSE_images} illustrates an application of the QRaFT package to a coronal image taken during the 2017-08-21 eclipse. The TSE image of the solar atmosphere exhibits a lot of structure at both small and large scales. The fine-scale azimuthal structure present in the enhanced image (Fig. \ref{fig_TSE_images}d) is much more detailed and geometrically consistent compared to the COR1 image, and its segmentation with the QRaFT code produces a large ensemble of features, many of which have successfully passed the validation conditions. The detected features yield a nearly uniform coverage of the observed TSE corona, making it possible to characterize the coronal geometry at most image locations (Fig. \ref{fig_TSE}). It can also be noticed that some of the visually identifiable quasi-radial structures are missed by the tracing code. These structures can be captured by further optimizing  QRaFT settings and/or running the code repeatedly with a broader range of processing keywords (Table. \ref{tab:keywords}). 

The presenting example illustrates how efficient the QRaFT segmentation can be when the quality of the studied coronal image is sufficiently high. There are reasons to expect that the new-generation operational coronagraphs should approach, and possibly exceed, this quality standard. At that point, segmentation results similar to those shown in Fig. \ref{fig_TSE} could be produced on a systematic basis, serving the needs of the computational solar physics community and improving the accuracy of space weather forecasts across the heliosphere.

\section{Conclusions}
\label{sec:conclusions}

We have presented an in-depth description of the mathematical algorithms used in the QRaFT image segmentation method, and demonstrated its performance on several types of coronal images. Processing synthetic images obtained from an MHD model made it possible to compare the image-extracted features with the ground-truth geometry of the coronal magnetic field in the model. The results of this comparison have shown that the typical angular misalignment between the QRaFT features and the POS magnetic field is about 4-7$^\circ$, with approximately 80\% of the traced locations showing a misalignment error of 10$^\circ$ or less. A more systematic performance assessment of the QRaFT method can be found in \citet{rura2025}.

The accuracy of the image tracing in the open-field coronal regions is found to be significantly higher than in the closed-field regions, which are not intended to be processed by the QRaFT algorithms. The performance of the method also reduces with the image SNR level. It is not a surprise that a high-quality TSE image has produced solid segmentation results, but it is significant that QRaFT has performed reasonably well when applied to a much lower-quality image, as illustrated by the STEREO COR1 example. As of now, we are not aware of another image segmentation code able to extract information about the magnetic field morphology from non-TSE observations of the weak and faint open solar corona. 

Our additional tests, which will be presented elsewhere, suggest that QRaFT performance can be further boosted by 
using more sophisticated image pre-processing techniques such as the temporal bandpass filtering developed by \citet{alzate2021}. Using machine learning algorithms for a more precise optimization of the processing keywords and a more accurate feature validation can further improve the tracing results. 

As stated earlier, image-based coronal segmentation techniques such as QRaFT can provide valuable information about the magnetic geometry of the corona for a wide variety of applications, including e.g. data-driven validation of global coronal models, verification of photospheric boundary conditions, quantification of the geometry and the topology of the coronal magnetic field based directly on observations, and an evaluation the accuracy and consistency of coronagraph products, to name a few. On a system-science level, QRaFT tracing results can be used to improve the performance of future and existing space weather simulation frameworks, such as e.g. ADAPT/WSA/ENLIL, by constraining the geometry of the simulated corona and benchmarking the accuracy of alternative solutions, and improving space weather forecasts. On the theoretical side, QRaFT analyses can contribute to resolving the open flux problem in the heliosphere, and provide new insights into the solar wind production mechanism and other aspects of the corona-heliosphere coupling.

%\begin{acknowledgements}
\bigskip

We are thankful to J. Klimchuk,  J. Davila, V. Troyan, and M. Aschwanden for useful discussions. V.M.U., C.E.R. and S.I.J. were supported through the Windows of the Universe Multi-Messenger program (NSF/AURA grant AST-0946422) and through the Partnership for Heliophysics and Space Environment Research (NASA grant No. 80NSSC21M0180). C.D. was supported by the NASA Living With a Star Science (80NSSC22K1021) and Living With a Star Strategic Capabilities (80NSSC22K0893) programs. N.A. acknowledges support from NASA ROSES through HGI grant No. 80NSSC20K1070.

%\end{acknowledgements}

%\bibliography{references}{}
%\bibliographystyle{aasjournal}

\appendix 
\label{sec:appendix}

\begin{deluxetable*}{l|lc}[h]
\tablecaption{Main subroutines of the QRaFT package \label{tab:subroutines}}
\tablewidth{0pt}
\tablehead{
\colhead{Name} & \colhead{Purpose} & \colhead{Equations} }
\decimalcolnumbers
\startdata
{QRaFT} & Top-level subroutine implementing all processing steps &  \ref{eq:trend} - \ref{eq:curv_cond} \\ 
read\_const\_string & Reads a vector of processing parameters from a formatted string & - \\
extract\_const & Reads a single processing parameter from a string & - \\
radial\_detrending & Removes radial intensity trend from image & \ref{eq:trend} - \ref{eq:detr} \\
rect\_to\_polar & Transforms image into polar coordinates & \ref{eq:polar} \\
get\_coordinates & Prepares a data structure used by rect\_to\_polar & - \\
patch\_image\_holes & Patches small groups of empty pixels using nearest-neighbor interpolation & - \\
smooth\_polar & Smoothens polar image using anisotropic kernel & \ref{eq:smooth} \\
azimuthal\_diff & Computes second-order azimuthal differences & \ref{eq:diff} \\
detrend\_azimuthal & Detrends polar image in azimuthal direction & \ref{eq:enh}\\
linspace & Creates a 1D array of uniformly spaced values & - \\
trace\_blobs & Identifies image blobs using adaptive thresholding & \ref{eq:thresh} \\ 
blob\_labeler & Creates an array of blob labels & \ref{eq:lbl} \\
blob\_analyzer & Computes geometric characteristics of blobs & \ref{eq:rho_set} - \ref{eq:phi_poly} \\
adapt\_thresh\_prob & Generates a set of percentile thresholds for an image & - \\ 
blob\_stat\_merger & Merges blob detection results obtained with different tracing parameters & \\
blob\_stat\_to\_features & Converts tracing results into a unified data structure compatible with QRaFT 2.0 & - \\
feature\_angles & Computed POS orientation angles of the detected features &  \ref{eq:feature_angles}-\ref{eq:feature_rad_angles} \\
feature\_validator & Returns a validated set of features satisfying filtering conditions & \ref{eq:phi_cond} - \ref{eq:curv_cond}\\
feature\_aggregator & Creates image-sized maps of orientation angles at each successfully traced location & - \\
%angles\_vs\_b & \makecell[l]{Computes misalignment angles \\ between images features and simulated magnetic field} & & & \ref{eq:h} - \ref{eq:b_angles} \\
angles\_vs\_b & Computes misalignment angles between images features and simulated magnetic field & \ref{eq:h} - \ref{eq:b_angles} \\
\enddata
\end{deluxetable*}

% const_str_ne = "px=10.88 cx=256 cy=256 dp=1 dr=2 sx=12 sp=3 sr=15 ps=2 pd=5 np=20 nr=10 p1=0.8 p2=0.99 r1=110 r2=0 w1=2 w2=10 l1=10 l2=30 vn=10 vi=0.2 vc=0.005"

%  const_str_COR1 = "px=10.88 cx=256 cy=256 dp=1 dr=2 sx=4 sp=3 sr=25 ps=2 pd=5 np=30 nr=10 p1=0.6 p2=0.99 r1=110 r2=0 w1=2 w2=10 l1=10 l2=30 vn=10 vi=0.2 vc=0.005"

%  const_str_TSE = 'px=4.77 cx=512 cy=512 dp=1 dr=2 sx=2 sp=2 sr=10 ps=2 pd=5 np=20 nr=10 p1=0.6 p2=0.99 r1=190 r2=0 w1=2 w2=10 l1=10 l2=60 vn=10 vi=0.02 vc=0.005'

\begin{deluxetable*}{llcccc}
\tablecaption{Processing keywords \label{tab:keywords}}
\tablewidth{0pt}
\tablehead{
\colhead{Name} & \colhead{Purpose} & \colhead{Units} & \colhead{PSI MAS} & \colhead{STEREO COR1} & \colhead{TSE}}
\decimalcolnumbers
\startdata
& {\bf  Original image } &  &  & & \\ 
   px & pixel size & arcsec & 10.88 & 10.88 & 4.77 \\
   cx & $x$-position of solar disk center & pixels & 256 & 256 & 512 \\
   cy & $y$-position of solar disk center & pixels & 256 & 256 & 512 \\ 
& {\bf  Coordinate transformation} &  & & & \\
   dp  & Azimuthal bin size & degrees & 1 & 1 & 1 \\
   dr  & Radial bin size & pixels & 2 & 2 & 2 \\
   r1  & Minimum processed heliocentric distance & pixels & 110 & 110 & 190 \\
   r2  & Maximum processed heliocentric distance  & pixels & 0 & 0 & 0 \\
&  {\bf  Smoothing and noise reduction} & & & & \\
   sx  & Size of Cartesian smoothing window  & pixels & 12 & 4 & 2 \\
   sp  & Azimuthal size of polar smoothing window & $\phi$ -bins & 3 & 3 & 2 \\
   sr  & Radial size of polar smoothing window & $\rho$ -bins & 15 & 25 & 10 \\   
& {\bf  Azimuthal differencing and detrending} & & & & \\  
   ps  & Scale of azimuthal differencing & $\phi$ -bins & 2 & 2 & 2 \\
   pd & Azimuthal scale used for detrending & $\phi$ -bins & 5 & 5 & 5 \\
& {\bf  Feature tracing} & & & & \\
   np  & Number of probability thresholds & $\in \mathbb{N} $ & 20 & 30 & 20 \\
   nr  & Number of inner boundaries & $\in \mathbb{N}$ & 10 & 10 & 10 \\
   p1  & Minimum percentile threshold & $\in (0,1)$ & 0.80 & 0.60 & 0.60 \\
   p2  & Maximum percentile threshold & $\in (0,1)$  & 0.99 & 0.99 & 0.99 \\
& {\bf  Feature validation }& & & & \\
   w1 & Minimum  width & $\phi$ -bins & 2 & 2 & 2 \\
   w2  & Maximum width & $\phi$ -bins & 10 & 10 & 10 \\
   l1  & Minimum length  & $\rho$ -bins & 10 & 10 & 10 \\
   l2  & Maximum length & $\rho$ -bins & 30 & 30 & 60 \\
   vn  & Minimum number of interpolation nodes & $\rho$ -bins & 10 & 10 & 10 \\ 
   vi  & Minimum intensity & $I_{enh}$ units & 0.20 & 0.20 & 0.02 \\
   vc  & Maximum feature curvature & $\in \mathbb{R}$ & 0.005 & 0.005 & 0.005 \\
\enddata
\end{deluxetable*}

\end{document}